\documentclass[iop]{emulateapj}

  \usepackage{graphicx}
  \usepackage[usenames,dvipsnames]{color}

  \newcommand{\kms} {~km~s$^{-1}$}
  
  \newcommand{\vlsr} {$v_{\rm LSR}$}
  \newcommand{\mo} {$M_{\odot}$}
  \newcommand{\lo} {$L_{\odot}$}
  \newcommand{\lbol}{$L_{\rm bol}$}
  \newcommand{\cmd} {~cm$^{-2}$}
  \newcommand{\cmt} {~cm$^{-3}$}
  \newcommand{\hcop} {HCO$^+$}
  \newcommand{\cdo} {C$^{18}$O}
  \newcommand{\tco} {$^{13}$CO}

  \newcommand{\iras} {IRAS~05327$+$3404} 
  \newcommand{\jyb} {Jy~beam$^{-1}$}

\begin{document}

\title{Millimetric and sub-millimetric observations of \iras\ ``Holoea'' in
  M36}

  \shorttitle{Mm and sub-mm observations of \iras}

\author{O. Morata}
  \affil{Institute of Astronomy and Astrophysics, Academia Sinica, P.O.\ Box
    23-141, Taipei 106, Taiwan}
  \affil{Department of Earth Sciences, National Taiwan Normal University, 88
    Sec.4, Ting Chou Rd., Taipei 116, Taiwan}
  \email{omorata@asiaa.sinica.edu.tw}
 
 \author{Y.-J. Kuan}
   \affil{Department of Earth Sciences, National Taiwan Normal University, 88
     Sec.4, Ting Chou Rd., Taipei 116, Taiwan}
 
 \author{P.~T.~P. Ho}
   \affil{Institute of Astronomy and Astrophysics, Academia Sinica, P.O.\ Box
     23-141, Taipei 106, Taiwan}
 
 \author{H.-C. Huang}
   \affil{Department of Earth Sciences, National Taiwan Normal University, 88
     Sec.4, Ting Chou Rd., Taipei 116, Taiwan}
 
 \author{E.~A. Magnier}
   \affil{Institute for Astronomy, University of Hawaii, 2680 Woodlawn Drive,
     Honolulu HI 96822}
 \and
 \author{R. Zhao-Geisler}
   \affil{Department of Earth Sciences, National Taiwan Normal University, 88
     Sec.4, Ting Chou Rd., Taipei 116, Taiwan}

  \shortauthors{Morata et al.}


\begin{abstract}
  The transition between the proto-star, Class~I, and the pre-main sequence
  star, Class~II, phases is still one of the most uncertain, and important,
  stages in the knowledge of the process of formation of an individual star,
  because it is the stage that determines the final mass of the star.
  We observed the YSO ``Holoea'', associated with \iras, which was classified
  as an object in transition between the Class~I and Class~II phases with
  several unusual properties, and appears to be surrounded by large amounts of
  circumstellar material.
  We used the SMA and BIMA telescopes at millimeter and sub-millimeter
  wavelengths to observe the dust continuum emission and the CO (1--0) and
  (2--1), \hcop\ (1--0) and (3--2), and HCN (1--0) transitions in the region
  around \iras.
  We detected two continuum emission peaks at 1.1-mm: SMM~1, the sub-mm
  counterpart of \iras, and SMM~2, $\sim6$ arcsec to the West.
  The emissions of the three molecules show marked differences.
  The CO emission near the systemic velocity is filtered out by the
  telescopes, and CO mostly traces the high-velocity gas.
  The \hcop\ and HCN emissions are more centrally concentrated around the
  central parts of the region, and show several intensity peaks coincident
  with the sub-mm continuum peaks.
  We identify two main molecular outflows: a bipolar outflow in an E--W
  direction that would be powered by SMM~1 and another one in a NE direction,
  which we associate with SMM~2.
  We propose that the SMM sources are probably Class~I objects, with SMM~1 in
  an earlier evolutionary stage.
\end{abstract}


\section{Introduction}

  The process of the formation of an individual star has been extensively
  studied and modeled in the last decades, which has resulted in a
  well-supported general picture \citep[see
    e.g.,][]{LadaWilking84,Adams87,Andre93} that describes the evolution of a
  young stellar object (YSO) from a pre-stellar core to a main-sequence star
  along several stages \citep{WardThompson96}.
  These steps can be mainly divided into two main phases.
  First, there would be an \textit{embedded phase}, in which a dense core in a
  molecular cloud collapses to form a heavily obscured, invisible to optical
  wavelengths, hydrostatic proto-star, surrounded by a progenitor disk and a
  massive envelope that proceeds to accrete the majority of its mass, while
  simultaneously driving a highly collimated bipolar outflow.
  This would be followed by a \textit{revealed phase}, after the accretion of
  material has stopped and once the YSO has acquired most of its final mass. 
  The enshrouding dust is progressively cleared away and a pre-main-sequence
  star, surrounded by a thinning disk, becomes visible to optical and
  near-infrared wavelengths.
  In time, this proto-star will complete its Kelvin-Helmholtz contraction onto
  the main sequence.

  A very interesting and still uncertain step in this picture is the
  transition between the proto-star and the pre-main-sequence star; between
  the embedded (Class I) and revealed (Class II) phases, which is accompanied
  by the action of energetic molecular outflows and ionized jets
  \citep{Magnier99b}.
  This stage approximately marks the end of the mass accretion onto the
  proto-star, which determines the final mass of the star.
  The mechanism that brings about this stop is not well understood yet, but it
  probably involves the action of the outflows on the surrounding dust and gas
  envelope.
  Thus, the observation of objects that happen to be found at this stage of
  the evolution is of fundamental importance to understand the physics of the
  processes involved and for the modeling of the disk and envelope structures
  around YSOs.
  In this paper, we present results of millimeter interferometric studies of
  one of these objects.

  The source nicknamed ``Holoea'', Hawaiian for ``flowing gas'', is a YSO
  associated with \iras.
  It was discovered in optical observations of the Galactic open cluster M36
  in Auriga by \citet{Magnier96}, although it is probably not associated with
  M36 and may be a distant member of the nearby S235 region.
  The distance to the object is then somewhat uncertain, ranging from the 1.2
  kpc adopted by \citet{Hron87} for M36 to the 1.6 kpc for S235
  \citep{Blitz82}.
  We adopt a distance of 1.2 kpc.
  \iras\ drives a powerful ionized outflow, seen in both CO (2--1) and optical
  spectra, with unusually high velocity ($\sim650$\kms) for a low-mass star.
  The structure of \iras\ is not clear yet \citep[see][]{Magnier99}.
  It could be a binary system, with a central star of an optical spectral type
  of K2 III, which is probably an FU Orionis star similar to L1551 IRS 5
  \citep{Magnier99}, and a still embedded young star that would be powering
  the outflow.

  The young star was classified \citep{Magnier99b,Magnier99} as a transitional
  YSO between Class I, because of the rising spectral energy distribution and
  a molecular bipolar outflow, and Class II YSOs, due to the visible central
  star and an ionized outflow.
  The spectral energy distribution (SED) also shows the presence of large
  amounts of circumstellar material, which according to optical and near-IR
  observations of the reflection nebula seems to be arranged in a disk with a
  relatively wide central hole of $\sim33\arcdeg$ opening angle.
  The ionized flow, the CO outflow, and the hole are all roughly aligned, and
  tilted by $\sim45\degr$ to our line of sight \citep{Magnier96}.
  Additionally, the brightness of the central star has increased $>1.5$ mag
  since the 1954 POSS plates.
  \citet{Magnier99} suggest that the source might be in the process of
  becoming exposed and hypothesized that the unusually wide SED can be due to
  the relative isolation of \iras, which allowed the formation of a large
  circumstellar disk with high angular momentum without being disrupted by
  external sources.
  Thus, this object seems to be a good candidate to study the transition from
  the Class I to the Class II phase in the evolution of YSOs.
  Subsequent 3.6-cm VLA observations by \citet{AngladaRodriguez02} detected a
  source, named VLA~2, inside the error ellipsoid of \iras, which coincides
  within $\sim1$ arcsec of the optical position of \citet{Magnier96}.
  H$_2$O maser emission surveys by \citet{Wouterloot93}, \citet{Codella95} and
  \citet{Sunada07} did not detect any maser emission near \iras.

  The structure of this paper is as follows:
  in Sect.~\ref{observations} we describe the SMA interferometer observations
  made at 1-mm and the BIMA interferometer observations carried out in the
  3-mm band; in Sect.~\ref{results} we present the results obtained of the
  261~GHz continuum emission and of the CO (2--1), \cdo\ (2--1),
  \hcop\ (2--1), CO (1--0), \hcop\ (1--0), and HCN (1--0) lines; and in
  Sect.~\ref{discussion} we discuss the results and propose a picture for the
  structure of the gas in the region.
  Sect.~\ref{conclusions} gives our conclusions.

\section{Observations}
 \label{observations}

\subsection{SMA observations}

 \begin{figure}

   \centering

   \includegraphics[width=\hsize]{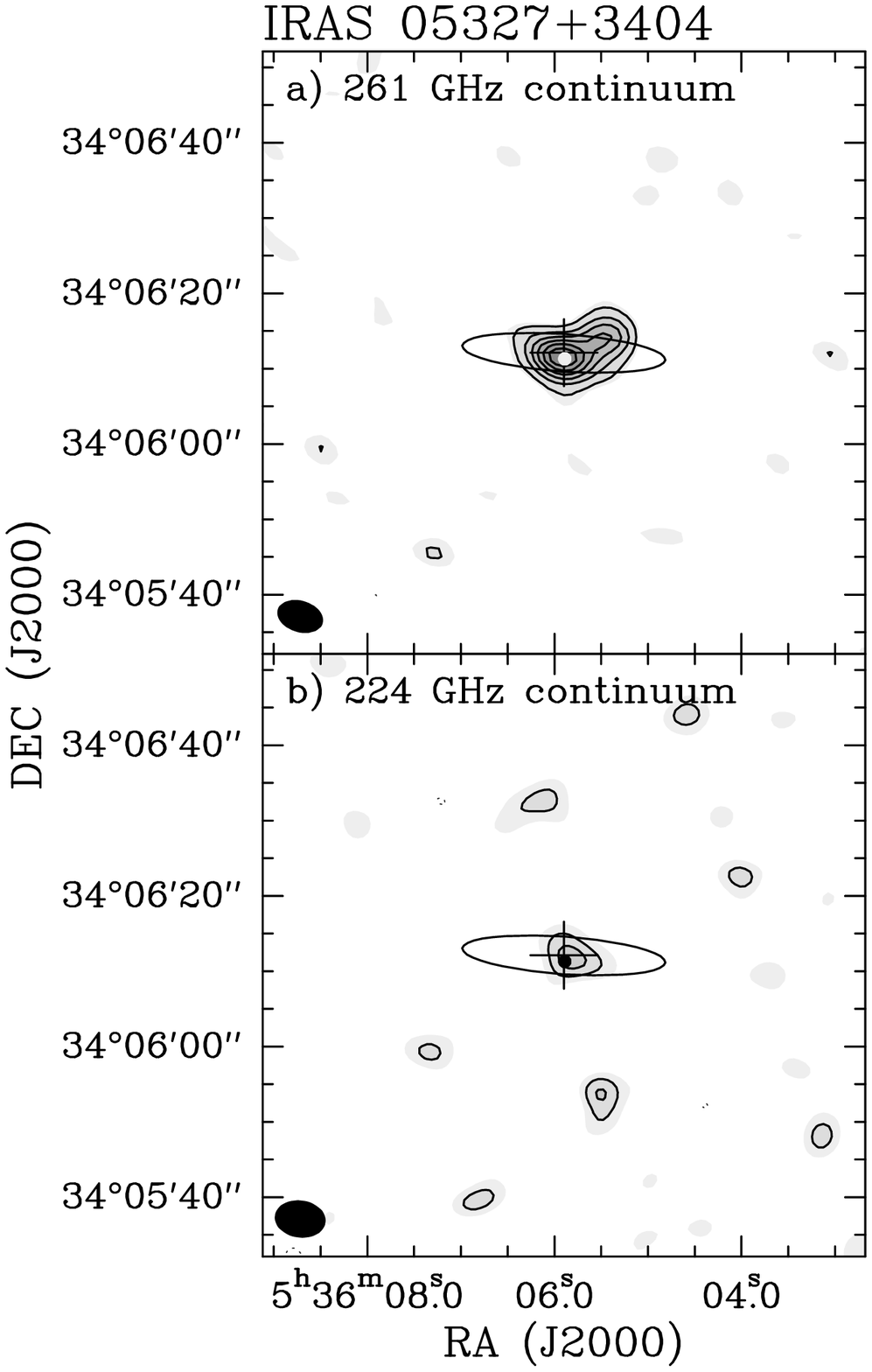}

   \caption{
     \textit{a)} Map of the 1.1-mm continuum emission in the \iras\ region.
     Contours are $-3$, 3, 6, 9, ... times the rms of the map, 1.35~m\jyb.
     The cross marks the position of ``Holoea'', \iras, and the filled (white)
     circle marks the position of the 3.6-cm VLA~2 source detected by
     \citet{AngladaRodriguez02}.
     The central ellipse marks the error ellipse of the IRAS source.
     The beam size is $\sim6\farcs2\times4\farcs2$, with a position angle,
     PA$=73.3\arcdeg$.
     \textit{b)} same as a) for the 224~GHz conti\-nuum after applying a
     \textit{uv} tapering of 36 k$\lambda$.
     Contours are $-3$, 3, 4 times the rms of the map, 2.8~m\jyb.
     The beam size is $\sim6\farcs8\times4\farcs9$, with PA$=82.0\arcdeg$.
   }

 \label{mapcontinuum}
 \end{figure}

 \begin{table*}

   \caption{
     Parameters of the 2D Gaussians fitted to the 261~GHz continuum map
   }

   \centering

   \addtolength{\tabcolsep}{0.35pt}

   \begin{tabular}{lcccccccccc}

     \hline\noalign{\smallskip}
      & & & Gaussian$^a$ & & I$_{peak}^b$ & Flux$^c$ & Size$^d$ &
       Mass$^e$ &  $N(\mathrm{H_2})\,$$^f$ & $n(\mathrm{H_2})\,$$^f$\\
      Source & R.A.~(J2000) & Dec.~(J2000) & size & P.~A. & (\jyb) & (Jy) &
       (arcsec) &  (\mo)& ($10^{23}$\cmd) & ($10^6$\cmt)  \\
      
     \noalign{\smallskip}\hline\noalign{\smallskip}
      SMM~1 & 5:36:05.93 & 34:06:11.5 & $4\farcs0\times2\farcs6$ &
        $+66\arcdeg8$ & 0.028 & 0.039 & 3.2 & 1.5 & 2.1 & 5.3\\
      SMM~2 & 5:36:05.45 & 34:06:14.2 & $2\farcs5\times2\farcs3$ &
        $-88\arcdeg6$ & 0.016 & 0.020 & 2.4 & 0.8 & 1.9 & 6.6 \\ 
     \noalign{\smallskip}\hline\noalign{\smallskip}

   \end{tabular}

   \begin{minipage}{18cm}

     \begin{list}{}{\leftmargin 1.5em \rightmargin 1em}

       \addtolength{\itemsep}{-2pt}

       \item[$^a$] Deconvolved major and minor axes.

       \item[$^b$] Peak intensity calculated from the fit of two 2D Gaussians
         to the continuum emission.

       \item[$^c$] Total integrated flux inside the 2D Gaussians resulting
         from the fit of the continuum emission.

       \item[$^d$] Effective diameter of the condensation from $\sqrt{\theta_a
         \theta_b}$, where $\theta_a$ and $\theta_b$ are the lengths of the
         major and minor axes of the Gaussian, respectively.

       \item[$^e$] Total mass calculated from the flux density integrated over
         the fitted 2D Gaussian, using the expression from \citet{Frau10}, for
         a distance of 1.2~kpc.

       \item[$^f$] $N(\mathrm{H_2})$ and $n(\mathrm{H_2})$ are the values of
         the column and volume densities averaged over the size of the fitted
         2D Gaussian, respectively, calculated using the expressions from
         \citet{Frau10}, assuming dust temperature, $T_{\rm dust}=15$~K, a
         standard dust-to-gas ratio of 100, a dust absorption coefficient at a
         frequency of 261 GHz, $\kappa_{261} = 0.0089$ cm$^2$ g$^{-1}$, taken
         as the value for dust grains with thin ice mantles for a volume
         density of $\sim5\times10^5$\cmt, \citet{Ossenkopf94}.

     \end{list}

   \end{minipage}

 \label{tcontinuumfits}
 \end{table*}

  The observations of \iras\ were performed with the eight-element SMA array
  \citep{hosma04} in Mauna Kea (Hawaii) as part of a filler-time project in
  November 2011 and January 2012 in the compact and sub-compact
  configurations, respectively.
  The half-power beam width of the 6-m antennas is 54 arcsec at 230~GHz, and
  41 arcsec at 261~GHz.
  Maps were obtained with the visibility data weighted by the associated
  system temperatures, using natural weighting.
  Baselines range from 6 to 53 k$\lambda$ and from 6 to 40 k$\lambda$,
  respectively.
  The resulting synthesized beam sizes are $\sim2\farcs5$--$6\farcs5$. 
  The total on-source observing time was of 0.99 h for the November 2011
  observations and 4.62 h for the January 2012 observations.
  The central position of the maps for both tracks was located at $\alpha
  (J2000) = 5^{\rm h}36^{\rm m}05\fs90$, $\delta (J2000) =
  +34\arcdeg06'12\farcs1$.

  The receivers were tuned to a rest frequency of 230.53797~GHz for the
  observations in November 2011 and to 267.55762~GHz for the observations in
  Janua\-ry 2012, Doppler tracked to a velocity V$_{\rm LSR}$ of $-21.7$\kms.
  In the 230.5 GHz observations, the correlator was configured to observe
  three spectral windows to include the CO (2--1), \tco\ (2--1), and
  \cdo\ (2--1) lines, with 203 kHz (0.26\kms) resolution; 3.94~GHz were
  dedicated to observe the 1.3 mm continuum in both the upper and lower
  sidebands.
  For the 267.5~GHz observations, the correlator was configured to observe 4
  spectral windows:
  three windows with 203~kHz (0.23\kms) spectral resolution, including the
  \hcop\ (3--2) line, and one window with 812~kHz (0.91\kms).
  3.94~GHz were dedicated to observe the 1.1 mm continuum, in both the upper
  and lower sidebands.

  The data were reduced using the MIR and MIRIAD \citep{Sault95} packages.
  The data were flagged for bad channels, antennas, weather, and pointing.
  We used 3C84 and BL~Lac as bandpass calibrators for the November 2011 track
  and 3C279, Mars, and 0927+390 for the January 2012 track.
  Gain calibrators were 0646+448 and 0530+135 for the November 2011 data and
  3C111 for January 2012.
  Absolute flux calibration was done using observations of Uranus in both
  tracks.

\subsection{BIMA observations}

  The observations of \iras\ were carried out with the BIMA array at the Hat
  Creek Radio observatory during 6 periods, using 6 antennas in the C
  configuration in August 1995 (twice) and September 1995, and using 9
  antennas in the H configuration in December 1995 and May 1996 (twice).
  Maps were done at 88 and 115 GHz with the visibility data weighted by the
  associated system temperatures, using natural weighting, and applying the
  primary beam correction.
  The $uv$-coverage was $\sim1.2$--37 k$\lambda$, and the resulting
  synthesized beam sizes are $\sim$6--10 arcsec.
  The half-power beam width of the BIMA antennas is 120 arcsec at 100 GHz.
  The central position of the maps was also located at $\alpha (J2000) =
  5^{\rm h}36^{\rm m}05\fs90$; $\delta (J2000) = +34\arcdeg06'12\farcs1$.

  Two frequency setups were used, centered at 88 and 115 GHz, respectively.
  The target molecular lines were CO (1--0) at 115.27120 GHz, \hcop\ (1--0) at
  88.18852 GHz, and HCN (1--0) at 88.63185 GHz.
  The digital correlator was configured to observe simultaneously several
  molecular line transitions in the upper side band (USB) at spectral
  resolutions of 97.7 and 390.6 kHz/channel, which correspond to velocity
  resolutions of 0.25 and 1.00 \kms\ for CO, and 0.32 and 1.28 \kms\ for
  \hcop\ and HCN.
  CO lines were observed in one period in August 1995 and in September
  1995.

  Calibration and data reduction were performed using the MIRIAD software
  package.
  The source 0530$+$135 was used as phase calibrator, while Mars, Saturn,
  3C273, and 0530$+$135 were used as band-pass calibrators.
  The molecular line transitions tuned to the lower side band (LSB) did not
  show any emission.

\section{Results}
 \label{results}

\subsection{Continuum observations}

\subsubsection{261 and 224 GHz continuum}
 \label{sma_cont}

  Figure~\ref{mapcontinuum}a shows the continuum map of IRAS $05327+3404$
  obtained with the SMA from the combination of the continuum data from the USB
  and LSB at a nominal frequency of 261~GHz.
  The only emission over 3$\sigma$ is located in the central 10 arcsec of the
  map.
  The emission is extended in an East--West direction with a slightly NW bend
  in the Western tip.
  The emission peak (SMM 1) is located at $\sim1\farcs1$ S of the catalog
  position of \iras\ with an intensity of 28.7~m\jyb\ ($\sim21\sigma$).
  There is a secondary emission peak (SMM 2) located at an offset,
  $\Delta\alpha$,$\Delta\delta =$ ($-5\farcs2$, $+3\farcs1$) from SMM~1, with
  an intensity of 17.4~m\jyb\ ($\sim13\sigma$).
  The separation between the two peaks is $\sim6\farcs1$, about $\sim7300$~AU
  at the assumed distance to the source, and it probably indicates the
  presence of two objects. 
  The total flux inside the half-intensity contour encompassing both emission
  peaks is 33.4 mJy. 

  We tried to see if the continuum emission could be des\-cribed by the
  superposition of two elliptical Gaussians centered around the position of
  the two emission peaks.
  Table~\ref{tcontinuumfits} shows the parameters of the two 2-D Gaussians
  that best reproduce the continuum emission over $3\sigma$:
  the one corresponding to SMM~1 is elongated in a NE-SW direction, while the
  one corresponding to SMM~2 is almost circular in shape.
  The central positions, which we adopt as the nominal positions of the sub-mm
  sources, are slightly displaced with respect to the visual
  identification we had made, but well inside our angular resolution.
  The total integrated fluxes we measure are 39.1 and 19.9 mJy, respectively.

  The catalog position of \iras\ is located $\sim0.8$ arcsec NW from SMM 1,
  while the unresolved 3.6-cm VLA~2 source detected by
  \citet{AngladaRodriguez02} is located $\sim0.5$ arcsec SW of SMM~1.
  The uncertainty in the determination of the VLA position is probably of the
  order of $\sim1$ arcsec, given the beam size and signal-to-noise ratio of
  the cm observations.
  We would also expect that if the unresolved VLA~2 source had any significant
  contribution from SMM~2, the centroid of the centimeter emission would be
  displaced $>1$ arcsec to the NW.
  Thus, given that the positions of the infrared and centimeter emissions
  coincide within $\sim1$ arcsec from SMM~1, we conclude that SMM~1 is the
  sub-mm counterpart of \iras\ and VLA~2.

  Figure~\ref{mapcontinuum}b shows the continuum map obtained by combining the
  USB and LSB continuum data in the 230.5 GHz track, with a resulting nominal
  frequency of 224~GHz, after applying a \textit{uv} tapering of 36
  k$\lambda$.
  We do not find any extended emission on the map obtained with natural
  weighting.
  After applying the tapering, we find a condensation $\sim7.5\times5.5$
  arcsec, coinciding with the position of SMM~1, with an intensity peak of
  12.7 m\jyb\ and an estimated total flux of 18.9 mJy (measured at the
  $2\sigma$ contour).

 \begin{figure}

   \centering

   \includegraphics[width=\hsize]{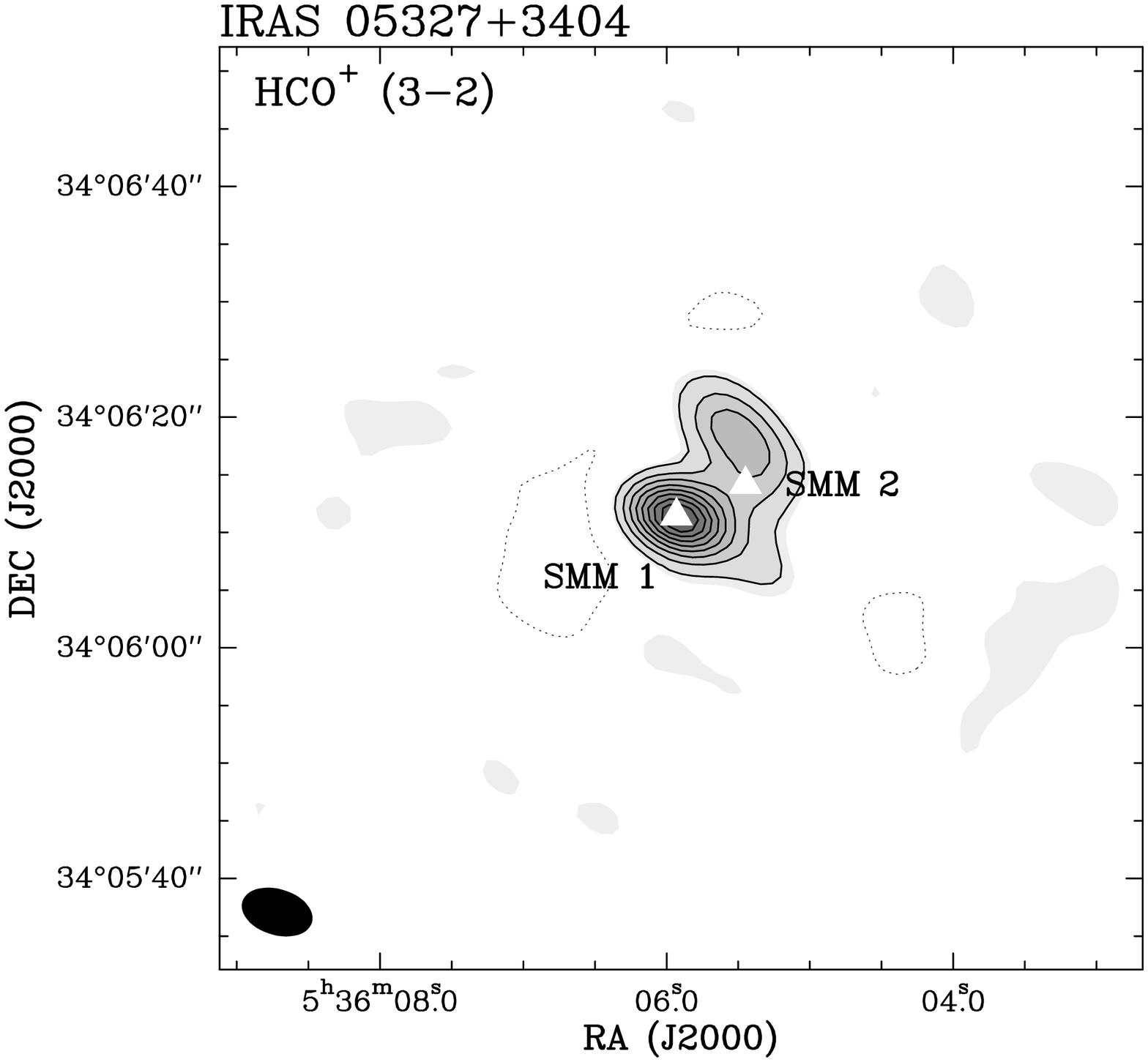}

   \caption{ 
     Integrated intensity map of the \hcop\ (3--2) emission in the velocity
     range from -23.99 to -19.21\kms, with a resulting mean velocity of
     $-21.6$\kms.
     Contours are $-3$, 3, 6, 9, ... times the rms of the map, 0.12 \jyb.
     The beam size is $\sim6\farcs3\times4\farcs0$, with PA$=73.3\arcdeg$.
     The triangles mark the positions of the emission peaks in the 261~GHz
     continuum map, SMM~1 and SMM~2 (Fig.~\ref{mapcontinuum}a).
   }

 \label{hcop32map}
 \end{figure}

 \begin{figure*}

   \centering

   \includegraphics[width=\hsize]{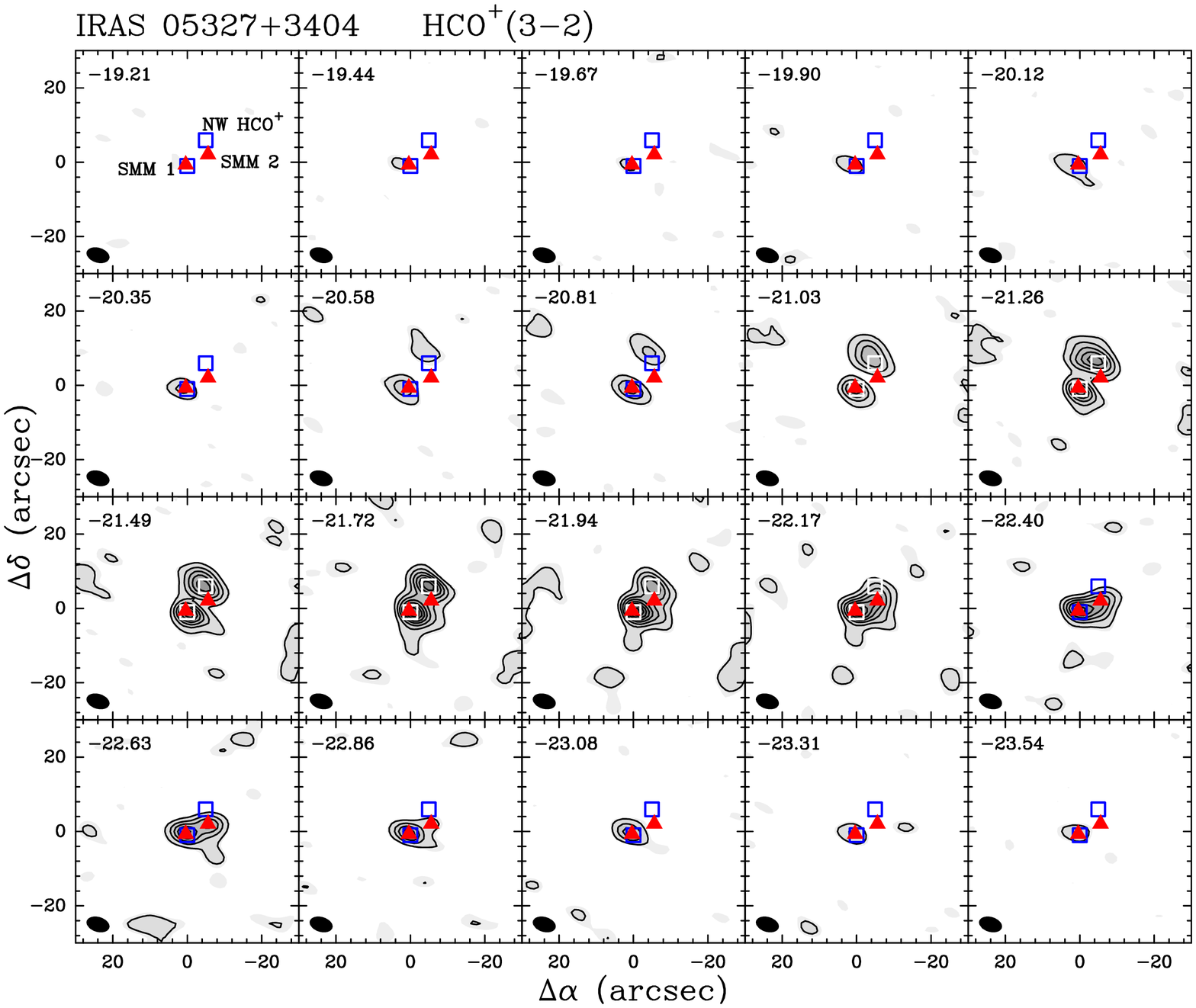}

   \caption{
     Channel maps of the \hcop\ (3--2) emission in the velocity range from
     $-23.54$ to $-19.21$~\kms.
     Contours are $-3$, 3, 6, 9, ... times the rms of the map, 0.26 \jyb.
     The beam size is $\sim6\farcs3\times4\farcs0$, with PA$=73.3\arcdeg$.
     The triangles mark the positions of the emission peaks in the 261~GHz
     continuum map (Fig.~\ref{mapcontinuum}a) and the squares mark the
     positions of the emission peaks of the integrated intensity map of
     \hcop\ (3--2).
     The velocity, in\kms, of each map is indicated in the upper left corner
     of each panel.
   }

 \label{hcop32chan}
 \end{figure*}

\subsubsection{3-mm continuum}

  The lines tuned up in the lower side band of both of the frequency setups
  used with the BIMA telescope, at 85 and 112 GHz, respectively, were too weak
  to be detected.
  The 768 channels available in four windows in the LSB could then be used to
  try to detect continuum emission.
  Unfortunately, the bandwidth was probably too narrow to detect any emission
  over 3$\sigma$ (1$\sigma$ is 3.8 and 20 m\jyb\ respectively) in the region.
  Adding the visibilities at the two frequencies in order to get a better
  signal-to-noise ratio using a multi-frequency cleaning did not improve the
  results and we failed to detect any emission over 3$\sigma$ (1$\sigma=4.9$
  m\jyb).

\subsection{Spectral observations}

\subsubsection{\hcop\ (3--2)}
 \label{hcop32}

  Figure~\ref{hcop32map} shows the integrated intensity map of the
  \hcop\ (3--2) emission obtained with the SMA.
  The emission is elongated in the NW-SE direction, with two partially
  resolved condensations.
  The more intense (peak intensity 3.38~\jyb) coincides with the position of
  the 1.1-mm continuum emission peak SMM~1.
  The second condensation (peak intensity 1.43~\jyb) is located $\sim8.5$
  arcsec NW of the emission peak.
  Both condensations are connected by low level emission at more than the
  3$\sigma$ level of the map.
  The continuum emission peak SMM~2 is located approximately at the neck of
  this bridge.

  Figure~\ref{hcop32chan} shows the channel maps of the \hcop\ (3--2) emission
  from $-23.8$ to $-19.2$\kms.
  The two condensations found in the integrated intensity map
  (Fig.~\ref{hcop32map}) are present in the central velocity channels.
  The central condensation is seen in all the channels containing any
  emission, while there is emission over a $3\sigma$ level $\sim7.5$ arcsec to
  the NW only from $-22.17$ to $-20.58$\kms.
  This emission coincides with the secondary condensation found in the
  integrated intensity map and seems to drift slightly from the SW to the NE.
  There is a third condensation only found in the more blue-shifted emission
  (from $-22.63$ to $-22.17$\kms) just south of the peak of the secondary
  condensation, located $\sim5.5$ arcsec to the NW of SMM~1, and coinciding
  with the position of SMM~2 (see Fig.~\ref{mapcontinuum}).
  All three condensations are connected by emission at levels $>6\sigma$ for
  velocities closer to the systemic velocity.
  We also find some weak extended emission elongated in a N--S direction just
  South of SMM~1 at the systemic velocity channel ($-21.72$\kms) and some
  towards the SW from a position in-between SMM~1 and SMM~2 at $-22.53$\kms.

\subsubsection{\hcop\ (1--0)}

 \begin{figure}

   \centering

   \includegraphics[width=\hsize]{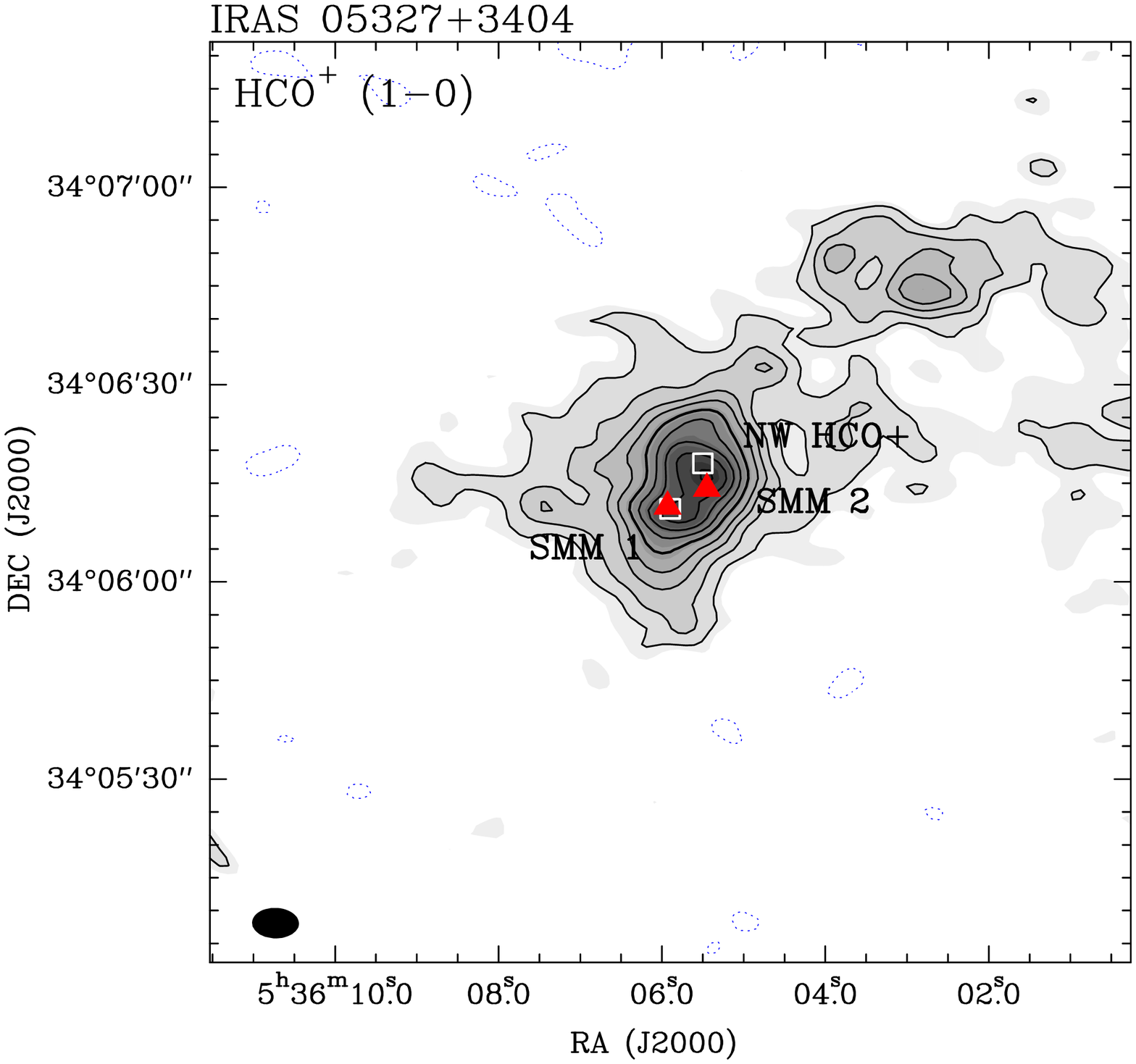}

   \caption{
     Map of the integrated intensity of the \hcop (1--0) emission in the
     velocity range from $-$22.42 to $-$20.78\kms, with a resulting mean
     velocity of $-$21.60\kms.
     The contours are $-3$, 3, 4, ..., 7, 9, ..., 15, 19, 21, 22, 23, 24 times
     the rms of the map, 0.060 \jyb.
     The triangles mark the position of the continuum peaks and the squares
     mark the position of the \hcop\ (3--2) emission peaks.
     The beam size is $\sim7\farcs1\times4\farcs6$, with PA$=88.6\arcdeg$.
   }

 \label{hcop-int}
 \end{figure}

 \begin{figure}

   \centering

   \includegraphics[width=\hsize]{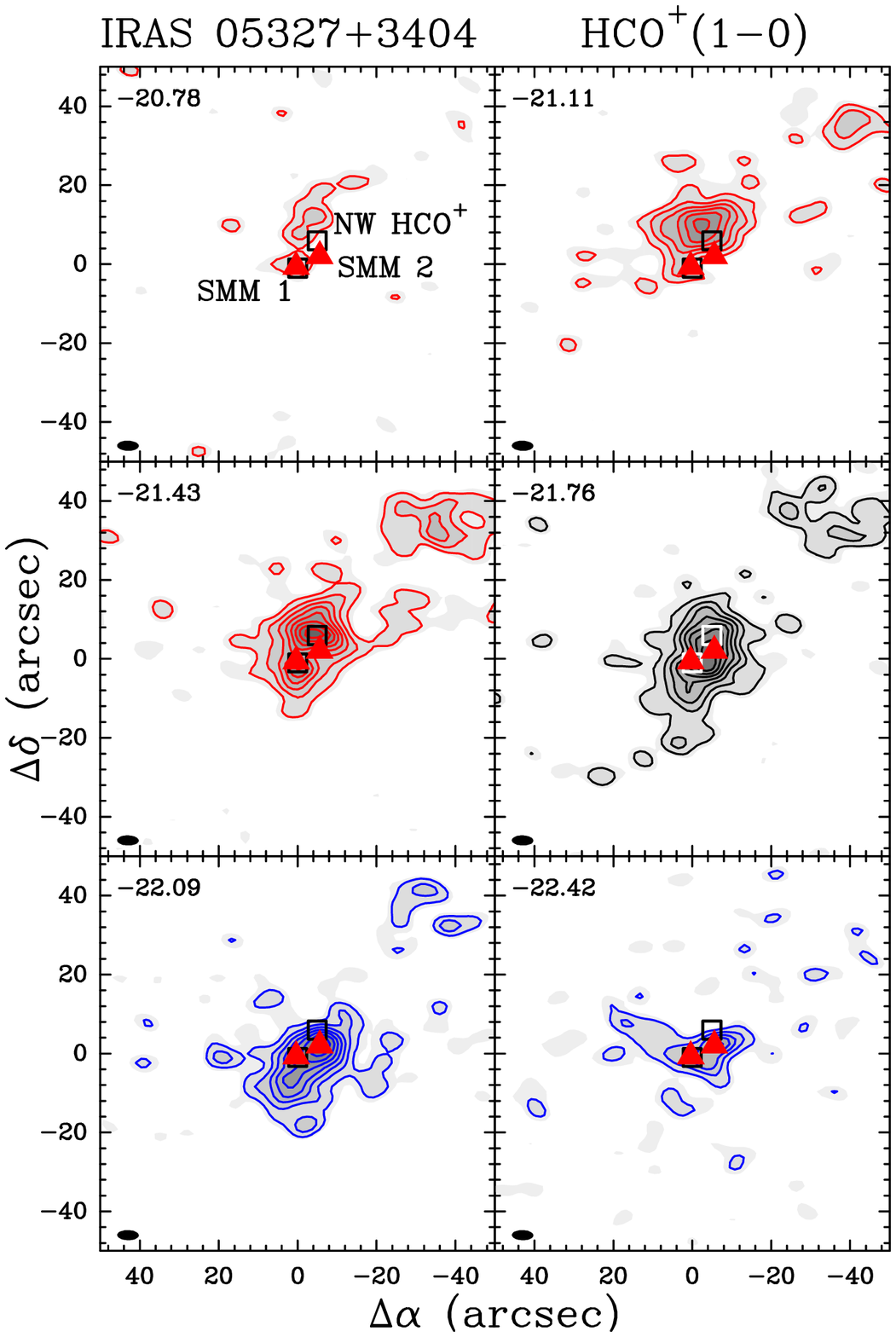}

   \caption{ 
     Channel maps of the \hcop\ (1--0) line in the velocity interval $-$22.42
     to $-$20.78\kms.
     The first contour is 0.447 \jyb\ with contour increments of 0.298 \jyb.
     The triangles mark the position of the continuum peaks and the squares
     mark the position of the \hcop\ (3--2) emission peaks.
     The peak intensity is found at \vlsr$=-21.43$\kms, at ($\Delta\alpha$,
     $\Delta\beta$) = ($-3\farcs5$,$+6''$), with an intensity of 2.7 \jyb.
     The velocity, in\kms, of each map is indicated in the upper left corner
     of each panel.
   }

 \label{hcop-channels}
 \end{figure}

 \begin{figure}

   \centering

   \includegraphics[width=\hsize]{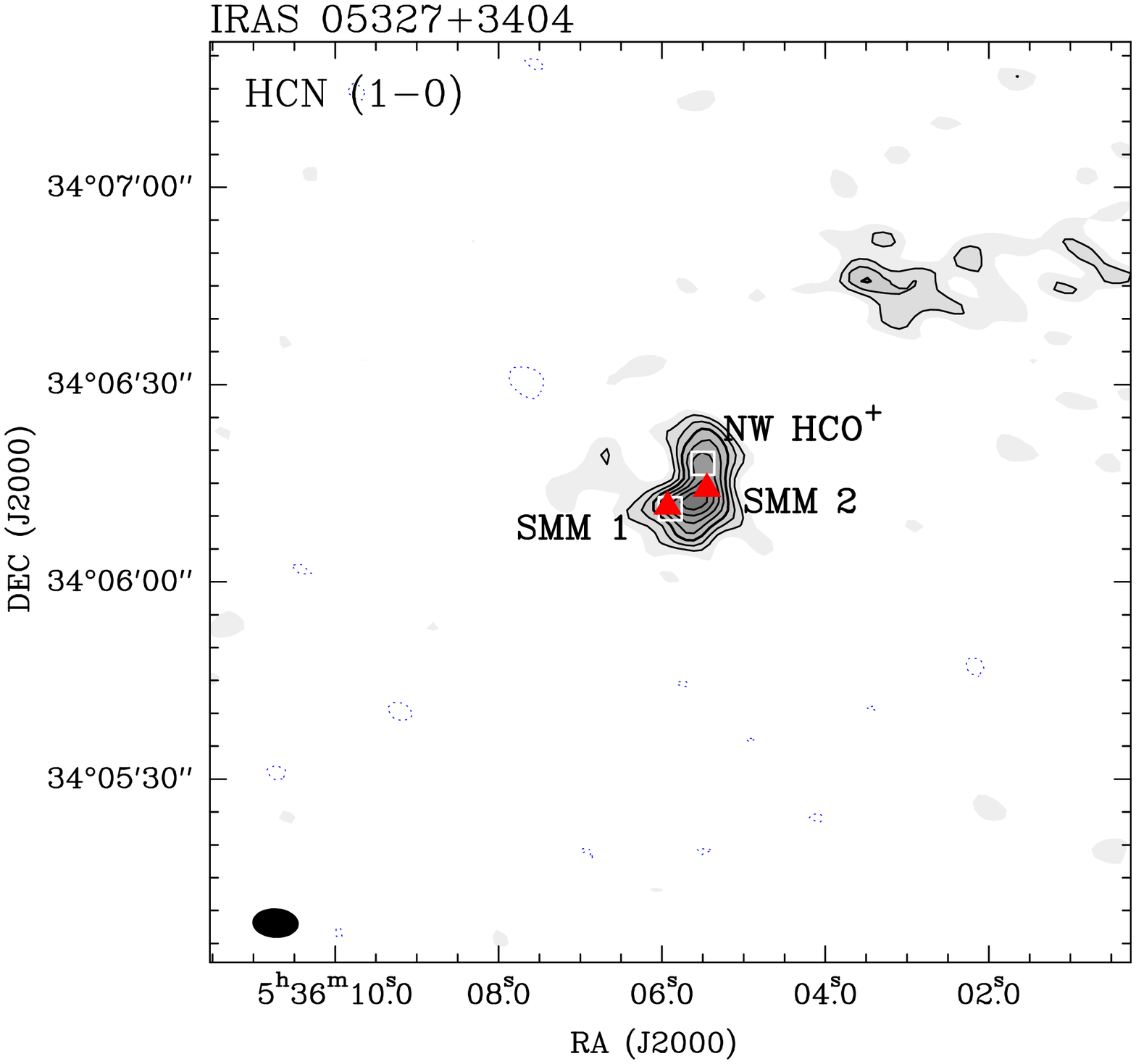}

   \caption{
     Map of the integrated intensity of the HCN (1--0) emission in the
     velocity range from $-$22.15 to $-$21.15\kms, with a resulting mean
     velocity of $-$21.65\kms.
     The contours are $-3$, 3, 4, ..., 9 times the rms of the map, 0.065 \jyb.
     The triangles mark the position of the continuum peaks and the squares
     mark the position of the \hcop\ (3--2) emission peaks.
     The beam size is $\sim7\farcs1\times4\farcs5$, with a PA$=87\arcdeg$.
   }

 \label{hcn-int}
 \end{figure}

 \begin{figure}

   \centering

   \includegraphics[width=\hsize]{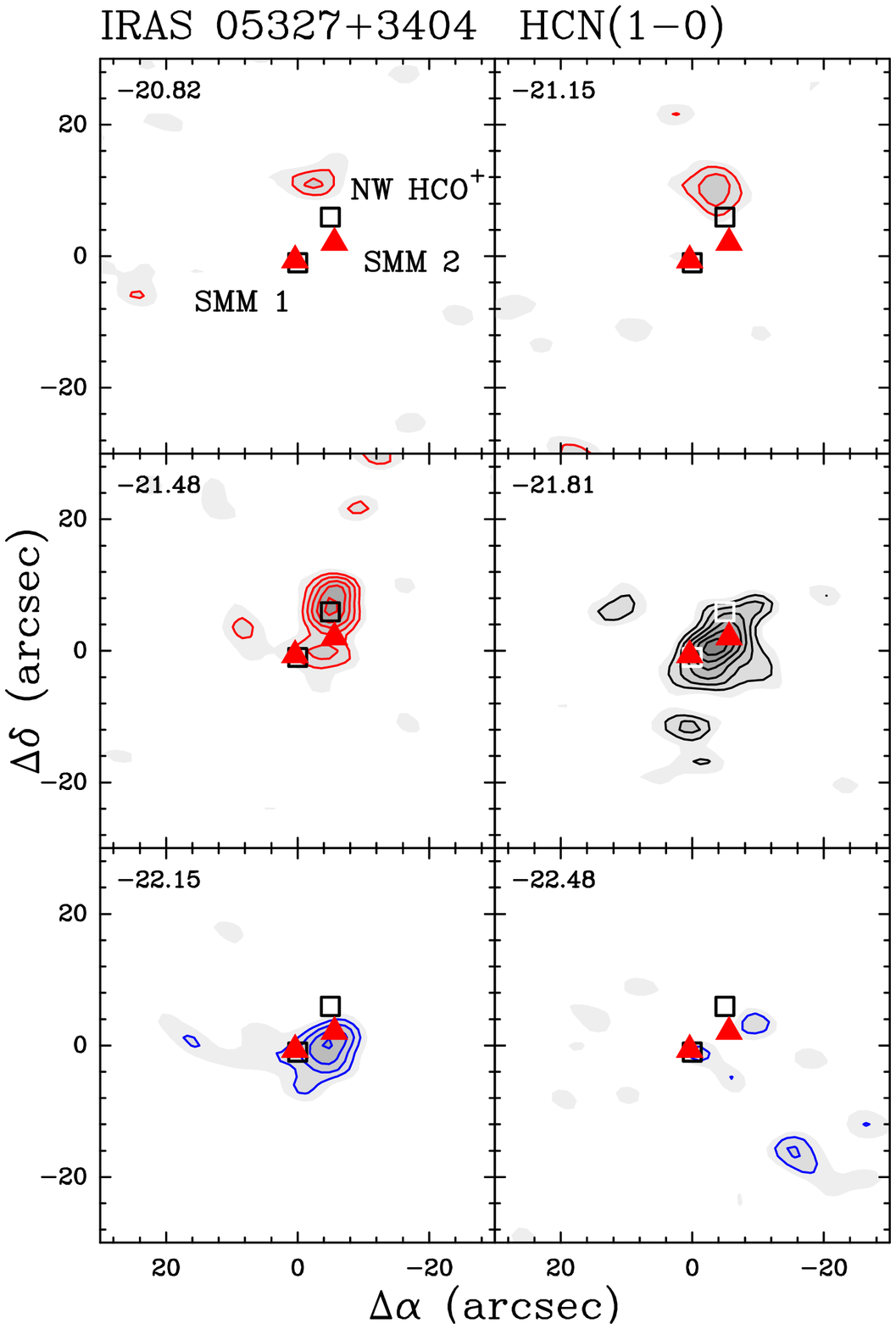}

   \caption{
     Channel map of the HCN (1--0) line in the velocity interval from $-$22.48
     to $-$20.82\kms.
     The contours are $-3$, 3, 4..., 7 times the rms of the maps, 0.129 \jyb.
     The triangles mark the position of the continuum peaks and the squares
     mark the position of the \hcop\ (3--2) emission peaks.
     The peak emission is found at $-21.81$\kms, at ($\Delta\alpha$,
     $\Delta\beta$) = ($-3''$,$0''$), with a flux density of 1.11 \jyb.
     The velocity, in\kms, of each map is indicated in the upper left corner
     of each panel.
   }

 \label{hcn-channels}
 \end{figure}

  Figure~\ref{hcop-int} shows the integrated intensity map of the emission of
  the \hcop\ (1--0) transition obtained with BIMA.
  The more intense emission is oriented in the NW-SE direction, enclosing the
  emission peaks found in the \hcop\ (3--2) maps.
  There are two emission peaks, the most intense one is located to the NW of
  SMM~1, at ($\Delta\alpha$, $\Delta\beta$) = ($-6\farcs4$,$+4\farcs2$), and
  it seems to include the two northwestern peaks found in the \hcop\ (3--2)
  maps.
  The second emission peak coincides within uncertainties with SMM~1.

  The channel maps of the \hcop\ (1--0) line (see Fig.~\ref{hcop-channels})
  also show the three condensations found in the (3--2) line, surrounded by
  more extended emission, which always connects the three main peaks.
  The central condensation is mainly found from $-22.42$ to $-21.43$\kms, and
  in particular around the systemic velocity.
  The emission towards the NW is predominantly found in the red-shifted
  velocities, from $-21.76$ to $-20.78$\kms.
  The emission around SMM~2 is particularly intense for the blue-shifted
  velocities, from $-22.42$ to $-21.76$\kms.
  There are other emission features in the blue-shifted channels, which were
  also seen in the \hcop\ (3--2) channel maps:
  an elongation in a N--S direction south of SMM~1, with a local emission
  peak; and an elongation to the SW from a point in-between SMM~1 and SMM~2.
  There is also a weak emission patch elongated to the NE of SMM~1 at
  $-22.42$\kms.

  There is also fainter emission at $\sim50$ arcsec NW of SMM~1, approximately
  coincident with emission also found in HCN (1--0) and CO (1--0) (see
  Sects.~\ref{emhcn} and \ref{emco}, respectively).
  This emission patch only appears in the velocity range $-22.09$ to
  $-21.11$\kms, but the emission peak moves $\sim10$ arcsec from W to E as the
  velocity changes from red- to blue-shifted channels.

\subsubsection{HCN (1--0)}
 \label{emhcn}

  The emission of the HCN (1--0) line obtained with the BIMA telescope
  (Fig.~\ref{hcn-int}) is very concentrated in the central part of the map, in
  the shape of an ``L'' that connects the two continuum peaks and the NW
  \hcop\ (3--2) emission peak.
  None of these three emission peaks coincides with the HCN emission peak,
  which is located $\sim4''$ to the NW of SMM~1, midway between SMM~1 and
  SMM~2.
  The NW \hcop\ (3--2) peak is seen as an emission plateau to the N.
  There is also faint emission at $\sim50$ arcsec NW of \iras, coinciding with
  the \hcop\ (1--0) emission patch.

  Figure~\ref{hcn-channels} shows the channel maps of the emission of the HCN
  (1--0) line in the range from $-22.48$ to $-20.82$\kms.
  The emission tends to be very concentrated and shows some similarities to
  the \hcop\ channel maps.
  The HCN (1-0) emission close to SMM~2 is predominantly red-shifted to the N,
  closer to the position of the NW \hcop\ (3--2) peak, and predominantly
  blue-shifted to the S of the source.
  Only the central channel maps show any, and weaker, emission around the
  position of SMM~1.
  The central velocity channel map also shows a weak condensation or local
  emission peak $\sim10$ arcsec South of SMM~1, approximately at the position
  of a similar condensation found in the channel maps of \hcop.

\subsubsection{CO (1--0)}
 \label{emco}

 \begin{figure}
 
   \centering

   \includegraphics[width=\hsize]{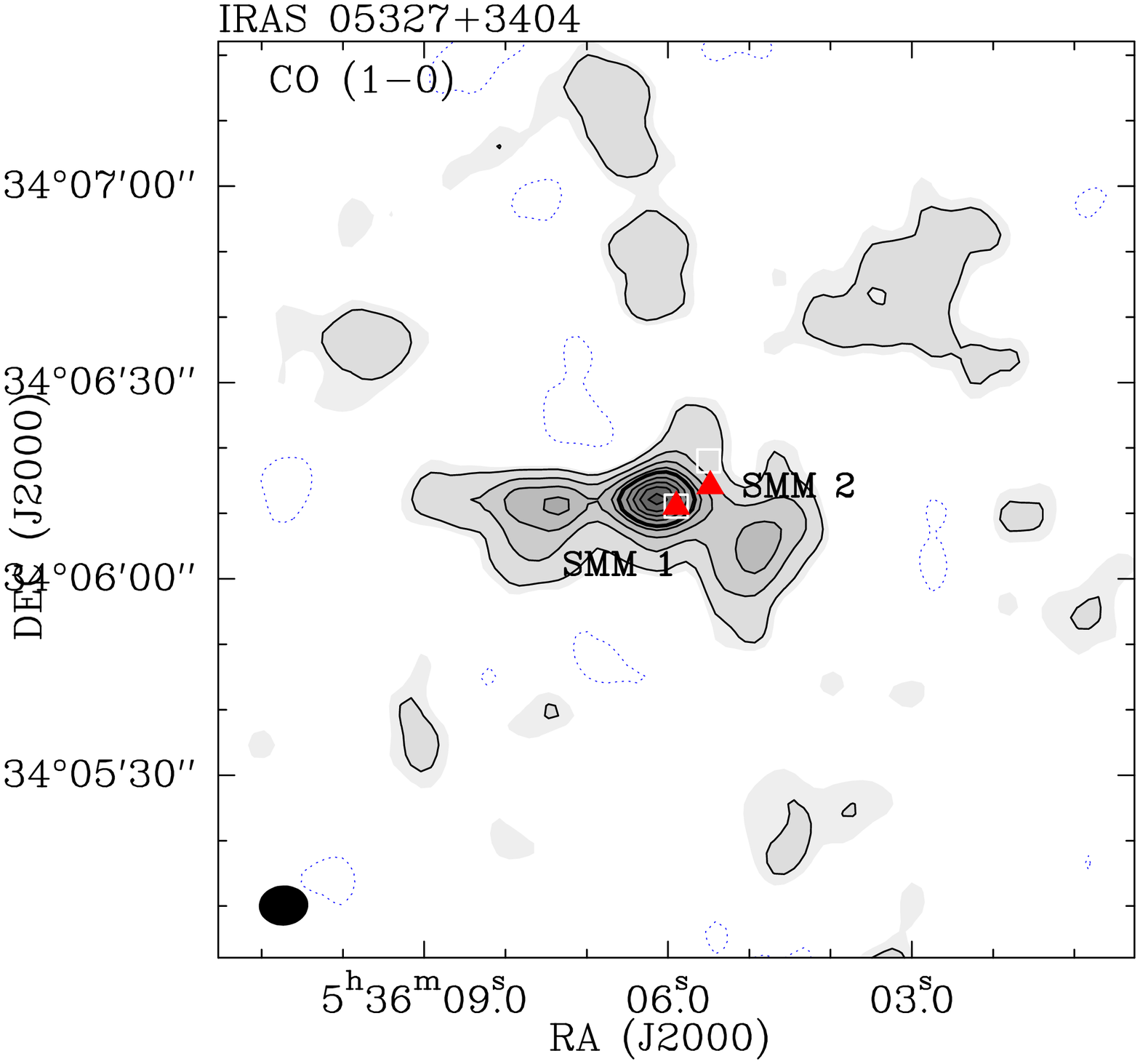}

   \caption{
     Integrated intensity of the CO (1--0) emission in the velocity range from
     $-27.47$ to $-15.04$ \kms, with a resulting mean velocity of $-21.01$
     \kms.
     The contours are $-3$, 3, 6, 9, ... times the rms of the map, 0.075 \jyb.
     The triangles mark the position of the continuum peaks and the squares
     mark the position of the \hcop\ (3--2) emission peaks.
     The beam-size is $\sim7\farcs.6\times6\farcs2$, with a PA$=-86.4\arcdeg$.
     The peak emission is located at a position ($\Delta\alpha$,
     $\Delta\beta$) = ($+2''$,$0''$) with an intensity of 2.14 \jyb.
   }
 
 \label{co-int}
 \end{figure}

  The map of the CO (1--0) emission obtained with BIMA (Fig.~\ref{co-int})
  shows a very different distribution from the maps of \hcop\ and HCN.  The CO
  emission is mainly found in an East-West strip passing towards the West
  through the position of SMM~1, and then bending in a SW direction West of
  SMM~1.
  The main emission peak is located $\sim3$ arcsec NE of SMM~1.
  There is a small spur in the emission in the NW direction from the position
  of \iras, which roughly corresponds to the more intense emission found in
  the \hcop\ (1--0) and HCN (1--0) integrated intensity maps.
  A weaker emission peak is found $\sim18$ arcsec East of SMM~1. 
  There is also a weak emission peak at $\sim15$ arcsec SW of SMM~1, in a
  similar direction as some weakly extended emission found in the
  \hcop\ channel maps.
  There are several clumps of emission in the NE direction from \iras, with a
  position angle (PA) $\sim9\arcdeg$.
  Another patch of emission is located $\sim50$ arcsec NW of \iras, coinciding
  with similar emission patches found in the \hcop\ (1--0) and HCN (1--0)
  maps.

  If we compare with the JCMT CO (2--1) integrated intensity map of
  \citet{Magnier99}, our observations cover the more central regions closer to
  \iras.
  Taking into account the lower angular resolution of the single-dish map, the
  distribution of the BIMA emission seems to be similar, but it is also clear
  that we are probably missing some extended emission.
  The CO (1--0) emission that we find along the NE direction follows the same
  direction as the NE outflow that was identified by \citet{Magnier99}.
  We are tracing the innermost part of this outflow and the clumps we identify
  coincide with local intense emission in the map of \citet{Magnier99}.

 \begin{figure*}

   \centering

   \includegraphics[width=\hsize]{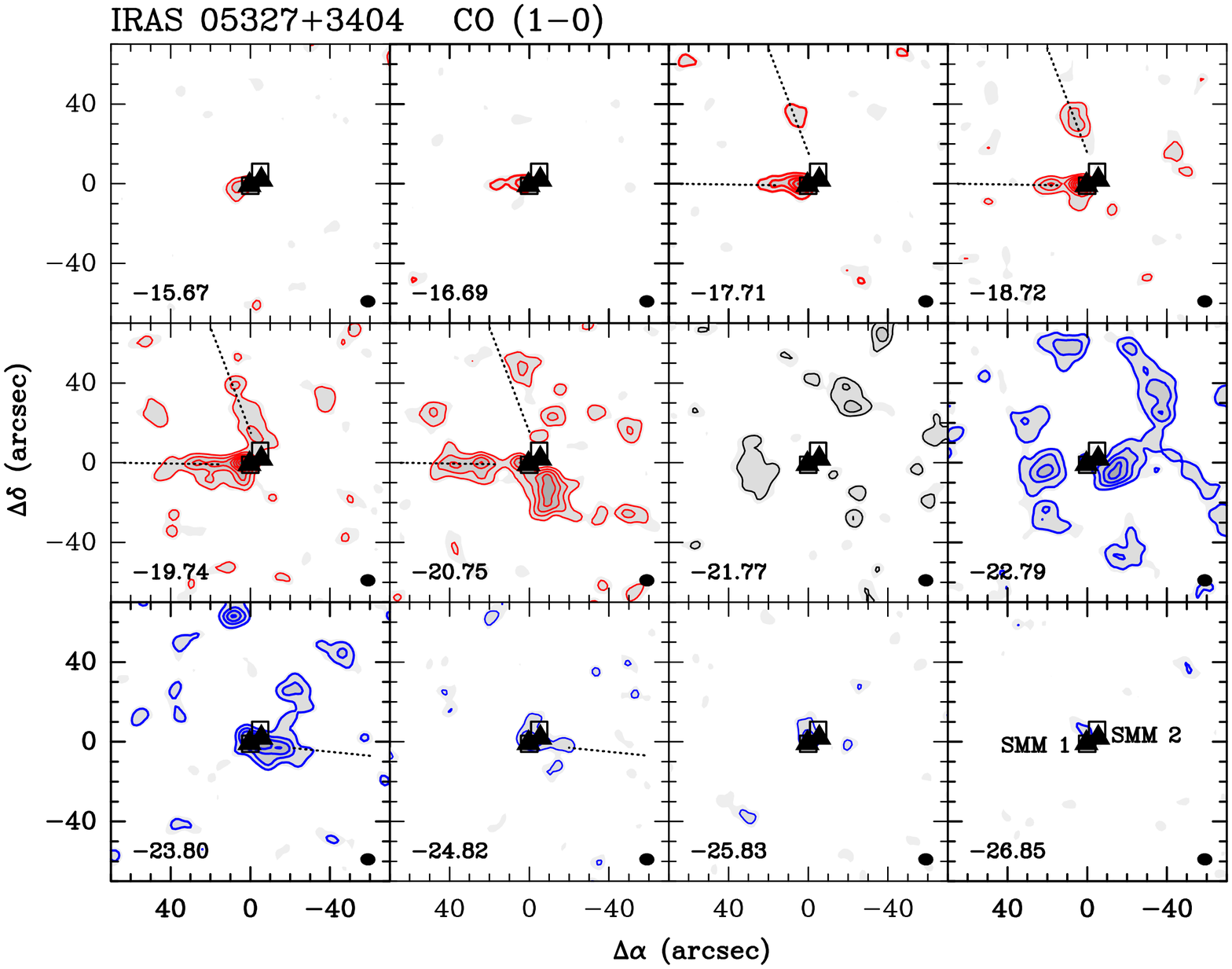}

   \caption{
     Maps of the 4-channel (0.25\kms/channel) average of the CO (1--0) lines
     in the velocity interval from $-$26.85 to $-15.67$\kms.
     The contours are 3, 6, 9,... times the rms of the map, 0.24 \jyb.
     The triangles mark the position of the continuum peaks and the squares
     mark the position of the \hcop\ (3--2) emission peaks.
     The peak emission is located at \vlsr$=-19.74$\kms\ at ($\Delta\alpha$,
     $\Delta\beta$) = ($+2''$,$0''$), with a flux density of 5.04 \jyb.
     The dotted straight lines mark the directions of possible outflow arms.
     The velocity, in\kms, of each map is indicated in the upper left corner
     of each panel.
   }

 \label{co-channel}
 \end{figure*}

\begin{figure}

   \centering

   \includegraphics[width=\hsize]{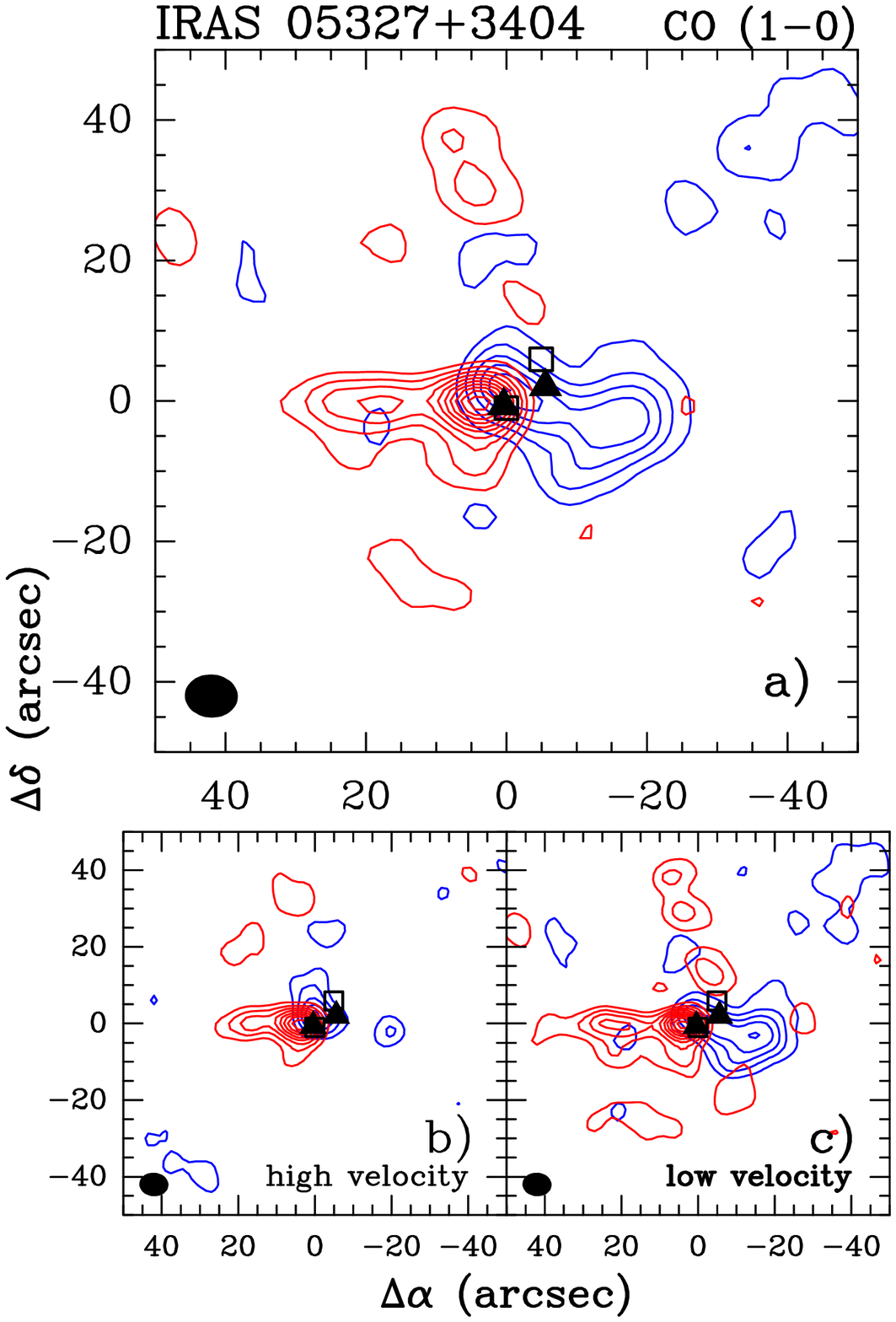}

   \caption{
     \textit{a)} Map overlapping the integrated emission of the red- (from
     $-20.63$ to $-15.80$ \kms) and blue-shifted (from $-26.22$ to $-22.66$
     \kms) channels of the CO (1--0) line in \iras.
     The lowest contour is 0.525 \jyb\ with increments of 0.35 \jyb.
     \textit{b)} Map overlapping the higher relative velocity emission of the
     red- and blue-shifted channels, from $-18.34$ to $-15.80$\kms\ and from
     $-26.72$ to $-24.95$\kms, respectively.
     The first contour is at 0.480 \jyb, with increments of 0.32 \jyb.
     \textit{c)} As panel \textit{b)} overlapping the lower relative velocity
     emission of the red- and blue-shifted channels, from $-20.63$ to
     $-18.60$\kms\ and from $-24.69$ to $-22.66$\kms, respectively.
     The first contour is at 0.750 \jyb, with increments of 0.50 \jyb.
   }

 \label{co-redblue}
 \end{figure}

  Figure~\ref{co-channel} shows the 4-channel (0.25 km s$^{-1}$/channel)
  averaged intensity maps of CO centered on the position of \iras.
  The channel at the systemic velocity, \vlsr$=-21.77$\kms, does not have any
  emission of CO in the central positions.
  This is probably due to an extended gas structure at this position at about
  the systemic velocity, which is filtered by the BIMA telescope (see
  Sect.~\ref{missingco}).
  On the other hand, the red- and blue-shifted emission channel maps reveal a
  very complex velocity structure.
  In order to examine more carefully the velocity structure of the region, we
  show in Fig.~\ref{co-redblue}a the superposition of the integrated intensity
  of the red- and blue-shifted channels with significant emission of CO and
  motions larger than 1\kms\ relative to the systemic velocity.
  We further show in Figs.~\ref{co-redblue}b and \ref{co-redblue}c the red-
  and blue-shifted integrated intensity maps for the high- and low-relative
  velocities with respect to the systemic velocity, respectively.

  Two outflows can be identified in the data: a bipolar outflow elongated in
  the East-West direction on both sides of SMM~1, accounting for most of the
  high-velocity emission in Fig.~\ref{co-redblue}, and another one in the NE
  direction, at an angle of $\sim9\arcdeg$, from which we seem to detect only
  the red-shifted lobe.
  These two outflows had also been detected in CO (2--1) by \citet{Magnier99},
  but at much lower angular resolution.
  We also find that the relative high-velocity emission is in general more
  compact than the relative low-velocity emission (see Figs.~\ref{co-redblue}b
  and \ref{co-redblue}c).

  The red-shifted ``arm'' of the East-West outflow extends eastwards up to
  $\sim45$ arcsec at low relative velocities, $\sim20$ arcsec for the
  high-relative velocities (see Figs.~\ref{co-channel} and \ref{co-redblue}).
  The blue-shifted ``arm'' is more clearly seen at low-relative velocities
  (Fig.~\ref{co-redblue}c) and it extends to $\sim25$ arcsec westwards.
  Figure ~\ref{co-redblue}c shows the overlap between the red- and
  blue-shifted lobes over the position of SMM~1.

  The NE outflow is more clearly seen at low-relative velocities (see
  Figs.~\ref{co-channel} and \ref{co-redblue}c).
  The maps at $-19.74$\kms\ in Fig.~\ref{co-channel} and
  Fig.~\ref{co-redblue}c also show that the outflow is not aligned with the
  position of SMM~1, but it is better aligned with the position of SMM~2, a
  few arcsec to the West of SMM~1.

  There are other velocity structures in our data that could be tracing more
  outflows, but they are not so clearly defined.
  Fig.~\ref{co-redblue}b shows red-shifted emission extending $\sim10$ arcsec
  to the SE of SMM~1, which is also found in the low-relative velocity
  emission map (Fig.~\ref{co-redblue}c).
  The high-relative velocity emission map also shows compact, $\sim10$ arcsec,
  blue-shifted emission elongated in a N--S direction originating from around
  SMM~1 (see Fig.~\ref{co-redblue}b).
  These red- and blue-shifted components could be roughly aligned, but it
  is very difficult to tell if they belong to independent outflows or are part
  of emission related to the two other outflows.
  Finally, the channel map at $-20.75$\kms\ shows a spur pointing away towards
  the SW from the central position of the map.
  The higher intensity contours point towards the position of SMM~2 and the NW
  \hcop\ peak, but the lower emission contours are connected to the position
  of SMM~1.

 \begin{figure}

   \centering

   \includegraphics[width=\hsize]{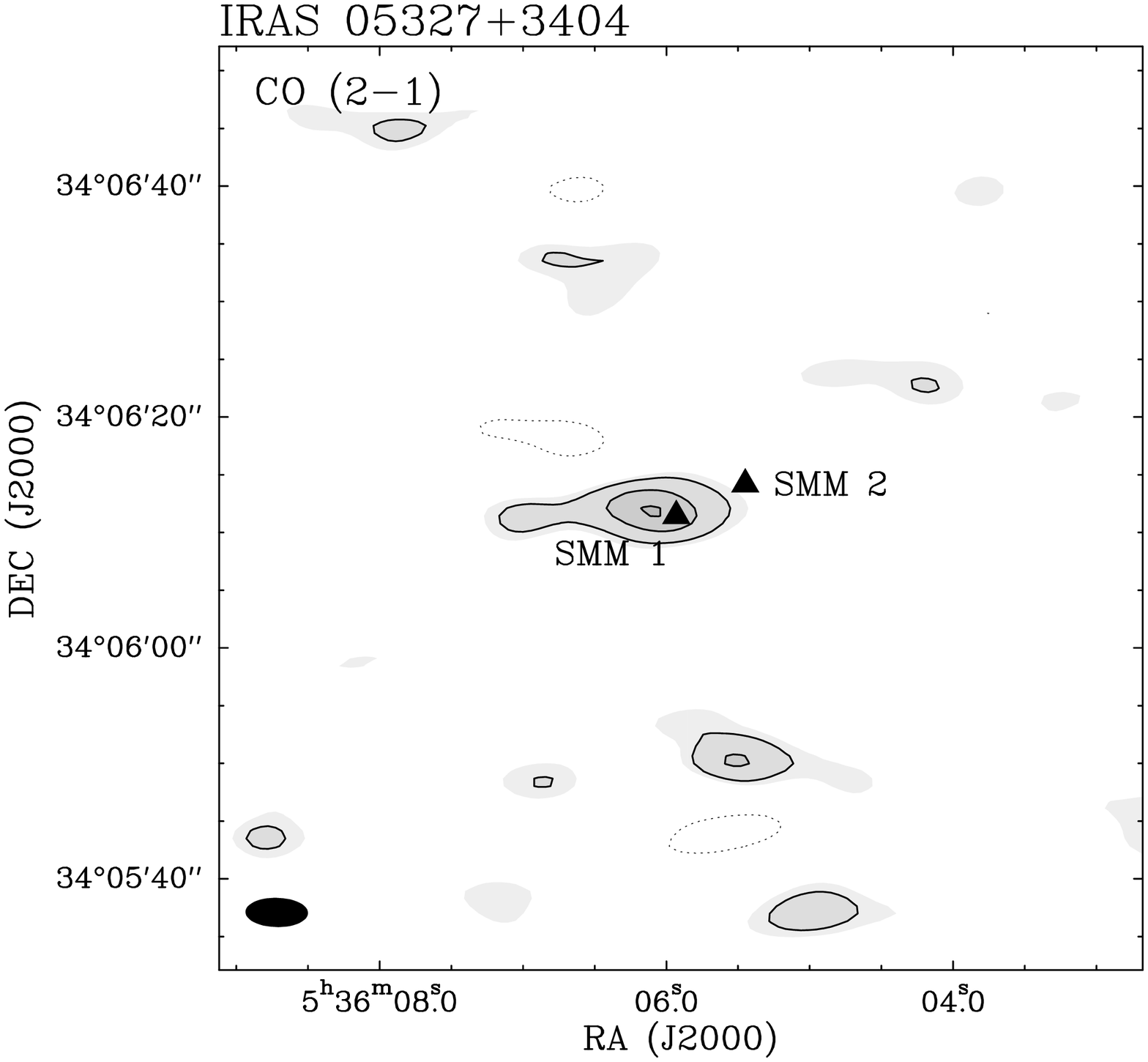}

   \caption{
     Integrated intensity map of the CO (2--1) emission in the velocity range
     from $-25.0$ to $-13.8$\kms, with a resulting mean velocity of
     $-19.4$\kms.
     Contours are -3, 3, 6, 9, ... times the rms of the map 0.19~\jyb.
     The beam size is $\sim5\farcs4\times2\farcs5$, with PA=88.8$\degr$.
     The triangles mark the positions of the emission peaks in the 261~GHz
     continuum map (Fig.~\ref{mapcontinuum}a).
   }

 \label{co21map}
 \end{figure}

 \begin{figure}

   \centering

   \includegraphics[width=\hsize]{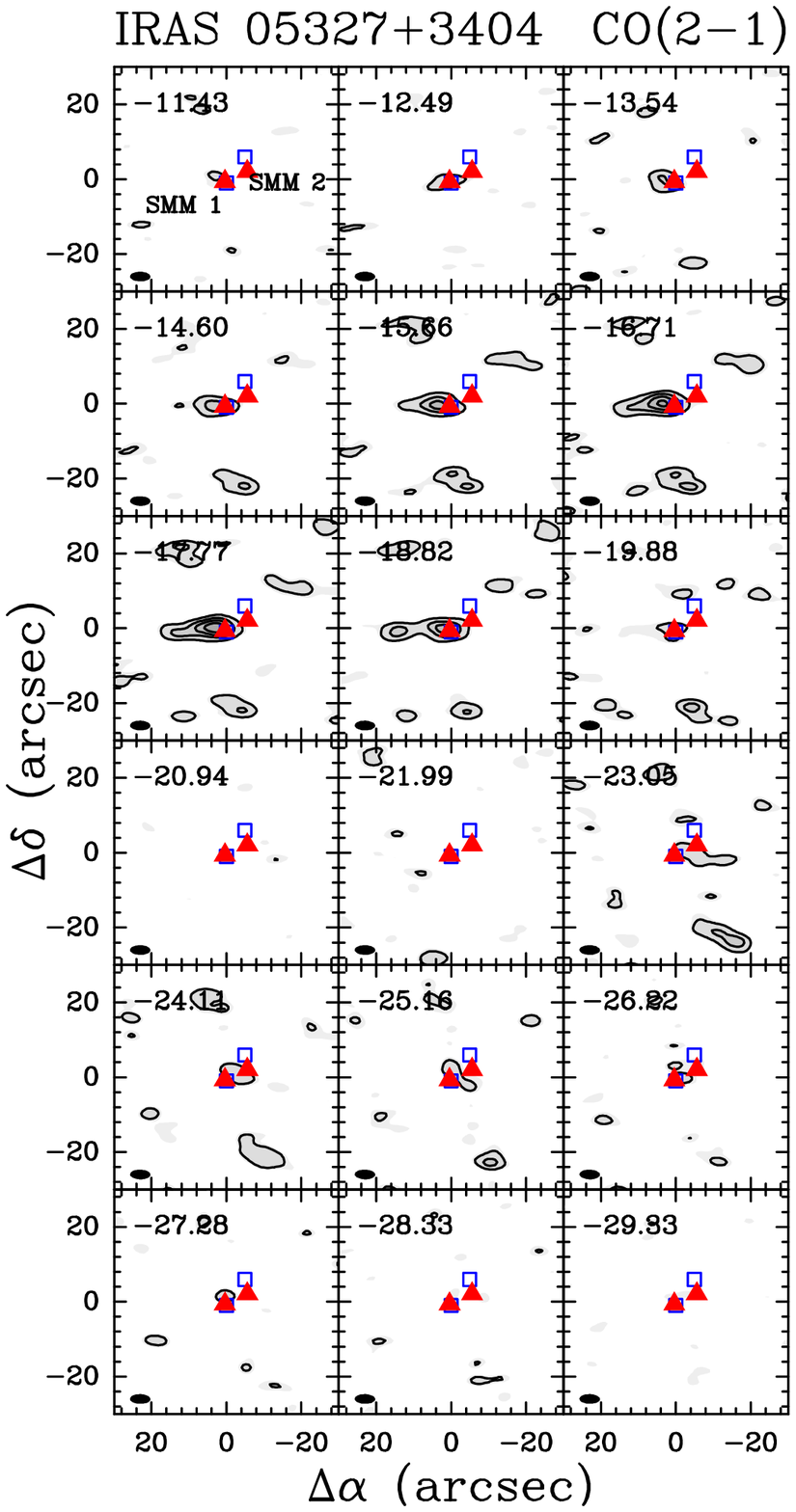}

   \caption{
     Maps of the 4-channel average of the CO (2--1) lines in the velocity
     interval from $-29.83$ to $-11.43$ \kms.
     The contours are 3, 6, 9,... times the rms of the map, 0.25~\jyb.
     The triangles mark the position of the continuum peaks and the squares
     mark the position of the \hcop\ (3--2) emission peaks.
     The velocity, in\kms, of each map is indicated in the upper left corner
     of each panel.
   }

 \label{co21-channel}
 \end{figure}

  Additionally, we find several patches of blue-shifted emission strewn around
  the central position, mostly in the NE and NW quarters at $-22.79$ and
  $-23.80$\kms.
  These emission patches probably trace more extended material partially
  filtered out by the telescope.

\subsubsection{CO (2--1)}

  Figure~\ref{co21map} shows the CO (2--1) integrated intensity map obtained
  with the SMA in the velocity interval $-25.0$ to $-13.8$\kms.
  The only significant emission is found in an E--W direction very close to
  SMM~1.
  The emission peak is located $\sim2.5$ arcsec West of SMM~1.
  There are not any other clearly detected structures in the map, possibly due
  to the limited observing time obtained with the SMA.

  Figure~\ref{co21-channel} shows the 4-channel averaged intensity maps of the
  CO (2--1) line.
  The emission is main\-ly found around the central position of the map, along
  the East--West direction through the position of \iras, in a way very
  similar to the one found in the CO (1--0) line:
  most of the emission is found in the red-shifted velocities in a East-West
  strip located to the West of \iras\ and originating from the position of
  SMM~1.
  The channels near the systemic velocity, $-21.76$\kms, do not have any
  emission.
  There is also some weaker emission in the blue-shifted maps more or less
  symmetrical to the emission found in the red-shifted maps.
  The red- and blue-shifted emissions overlap on the position of SMM~1, which
  supports the identification of SMM~1 as the powering source of the E--W
  outflow.

\subsection{Spectra}

\begin{figure}

   \centering

   \includegraphics[width=\hsize]{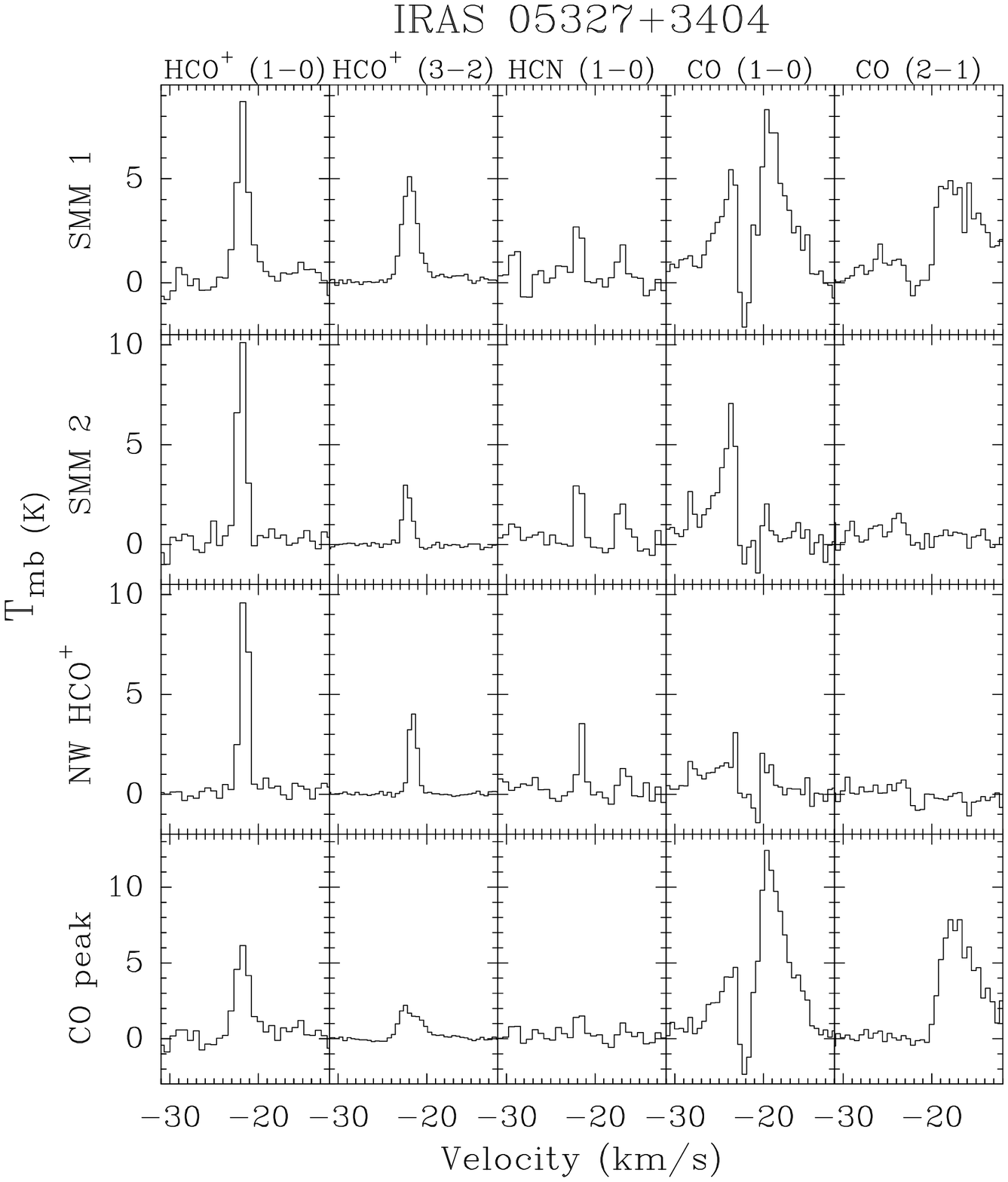}

   \caption{
     Spectra of the observed \hcop\ (1--0), \hcop\ (3--2), HCN (1--0), CO
     (1--0), and CO (2--1) lines obtained at four selected points of our maps.
   }

 \label{f-specmolecules}
 \end{figure}

  Figure \ref{f-specmolecules} shows the spectra of the five molecular lines
  we detected (\hcop\ 1--0, \hcop\ 2--1, HCN 1--0, CO 1--0, and CO 2--1)
  obtained at four selected positions of our maps:
  the sub-millimeter peaks, SMM~1 and SMM~2, the NW \hcop\ peak, and the CO
  peak. 
  We can appreciate the absorption dip in the CO emissions at the systemic
  velocities, where the \hcop\ and HCN emissions are found. 
  We can also see the broadening of the \hcop\ (3--2) line at the SMM~1
  position and, especially, at the position of the CO peak (see
  Sect.~\ref{kinemhcop}).
  Line-widths for HCN are $\sim0.7-1.1$\kms, while the line-widths for
  \hcop\ are $\sim1.0$--1.1\kms\ at the SMM~2 and NW \hcop\ positions, and
  $\sim1.4$--2.0\kms at the positions of SMM~1 and the CO peak (see
  Sect.~\ref{kinemhcop}).

\section{Discussion}
 \label{discussion}

\subsection{Dust continuum emission}
 \label{dustemission}

  We calculated the mass of the molecular gas from the 1.1-mm continuum
  emission, using the flux density and emission sizes measured from fitting
  two 2D-Gaussians to the continuum map.
  Table~\ref{tcontinuumfits} gives the H$_2$ mass, and H$_2$ averaged column and
  volume densities for the two condensations found in the 261~GHz continuum
  map.
  Given the presence of stellar objects embedded in the dust emission, we
  assumed a dust temperature of 15~K and a dust absorption coefficient for
  dust with thin ice mantles at 261~GHz, $\kappa_{261} = 0.0089$ cm$^2$
  g$^{-1}$ \citep{Ossenkopf94}.
  We estimate a mass of 1.5~\mo\ for SMM~1 and 0.8~\mo\ for SMM~2.
  The uncertainties in the determination of the dust temperature represent an
  uncertainty in the mass calculation of about a factor of 2.
  For dust temperatures of 10--20 K, the estimated masses would be between
  2.9--1.0~\mo\ and 1.5--0.5~\mo, respectively.
  These estimates assume a very simple spherical structure for the envelope of
  these sources.
  The real  distribution of the gas  around the SMM  sources probably includes
  the  presence of  circumstellar disks,  and  the mass  in the  proto-stellar
  objects, which would add some uncertainties to the mass determination.

  \citet{Magnier96} gave a range of values for the bolometric luminosity of
  \iras, \lbol$\sim$41--82~\lo.
  As we have shown in Fig.~\ref{mapcontinuum}, the IRAS error ellipse includes
  both SMM~1 and SMM~2, and it is conceivable that this bolometric luminosity
  has contribution from both sources.
  In the following discussion, we assume that most of the \lbol\ is associated
  with SMM~1, while \lbol\ for SMM~2 is not easy to define.

  In order to see if we could shed more light on the evolutionary stages of
  the SMM sources, we have compared the values of the mass from the sub-mm
  dust continuum with the envelope mass vs. bolometric luminosity plot shown
  in Fig.~1 of \citet{Bontemps96}.
  We define as envelope mass, $M_{\rm env}$, the mass derived from our sub-mm
  dust emission observations.
  Interestingly, SMM~1 would be located in the region mainly occupied by the
  Class~0 sources, while SMM~2 would be at the border between the regions
  where Class~0 and Class~I are found.
  The uncertainty in the determination of the envelope mass is more important
  to classify SMM~2 than the uncertainty in the determination of its
  bolometric luminosity (which we expect to be lower than that of SMM~1):
  a lower value for the calculated envelope mass will set SMM~2 deeper into
  the Class~I region.

\subsection{The molecular outflows: CO emission}
 \label{cokinem}

  The results of the CO (1--0) data shown in Sect.~\ref{emco} strongly
  indicate that there are several outflows in the region around \iras, with
  probably different inclinations with respect to the plane of the sky, and,
  correspondingly, there should be several objects powering them.
  To summarize, we have clearly identified two molecular outflows:
  one in an East-West direction, around SMM~1, which seems to be the more
  promising candidate to power this outflow; and another outflow elongated in
  a NE direction, with a PA$=9\arcdeg$, aligned with SMM~2, which is also the
  best candidate to power the outflow.
  Additionally, there are indications of other kinematic structures, but our
  data are not complete enough to clearly identify the nature of these
  structures:
  the ``SW arm'' found in the CO (1--0) data; the compact blue-shifted
  emission in a N--S direction North of SMM~1; and the compact red-shifted
  emission SE of SMM~1.

\begin{figure}

   \centering

   \includegraphics[width=\hsize]{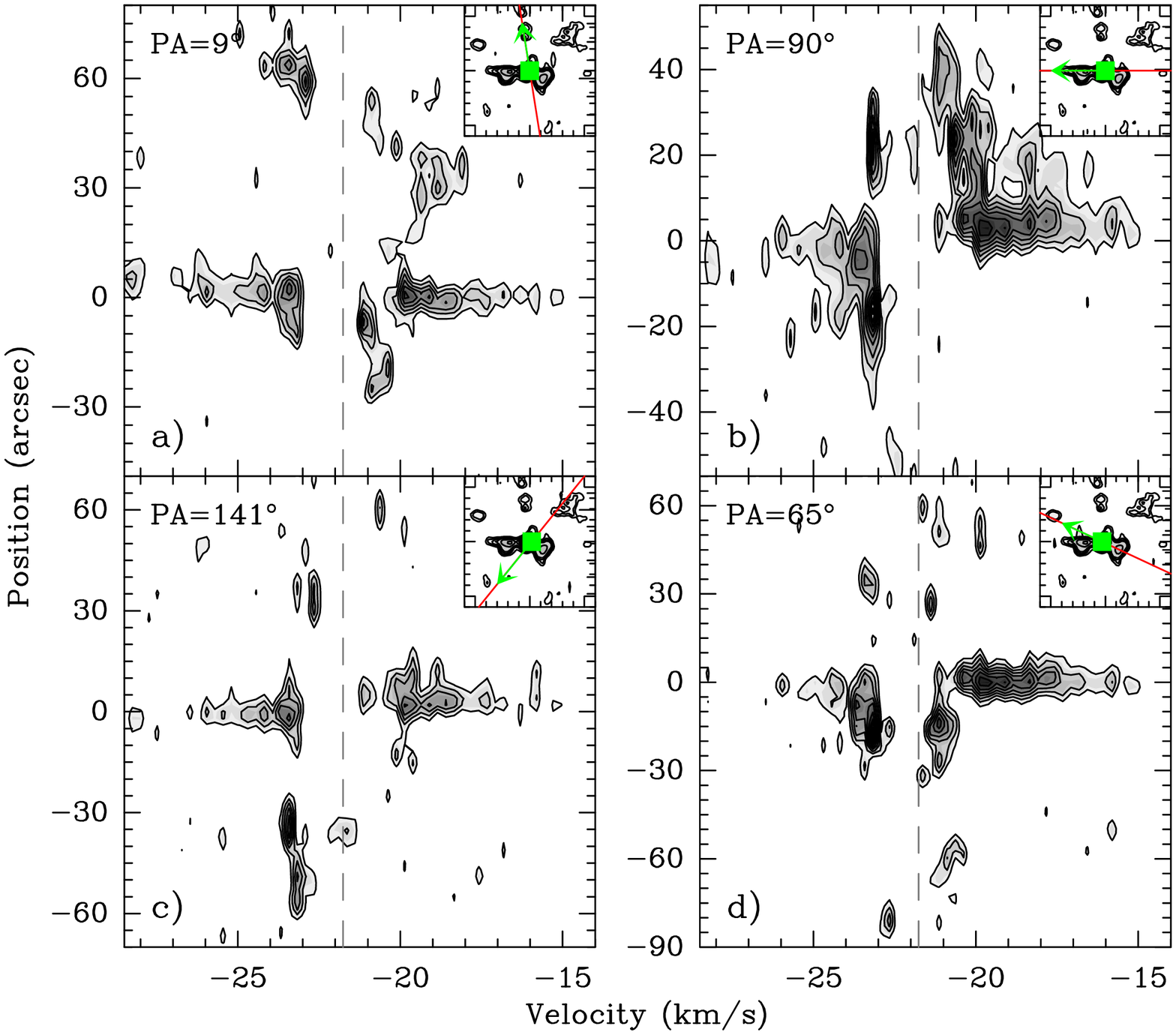}

   \caption{
     CO Position-Velocity cuts along four different position angles:
     \textit{top left panel a)} PA$=9\degr$, \textit{top right panel b)}
     PA$=90\degr$, \textit{bottom left panel c)} PA$=141\degr$, \textit{bottom
       right panel d)} PA$=65\degr$.
     The initial contour is at 1.19 \jyb, with increments of 0.79 \jyb.
     The dashed vertical line shows the position of the systemic velocity.
     The inset map for each figure shows the origin \textit{(green square)}
     and direction of the P-V cut \textit{(green arrow over the red line)}
     over the integrated intensity map of the corresponding transition.
     The position angle of the P-V cut is calculated with respect to the N
     direction.
     As a reference, positive positions correspond to the NE for
     PA$=9\arcdeg$ and $65\arcdeg$, E for PA$=90\arcdeg$, and SE for
     PA$=141\arcdeg$.
   }

 \label{co-pvcuts}
 \end{figure}

  Figure~\ref{co-pvcuts} shows the P-V cuts we took along the red- and
  blue-shifted elongated structures.
  The missing CO emission in the central regions of the map is very apparent
  in the velocity range $\sim-22.9$ to $\sim -21.4$\kms, which makes it very
  difficult to analyze what is really happening near \iras.
  Fig.~\ref{co-pvcuts}a shows the P-V cut in the direction of the optical
  outflow, PA$=9\arcdeg$.
  Similarly to what \citet{Magnier99} found in their data, the outflow
  velocity increases in relative velocity with the distance from the central
  region. 
  But, we also find that at larger distances the velocity decreases with
  distance.
  On the opposite side of the central region, there is a small gradient in the
  red-shifted velocities that probably points to the SW ``outflow'' found in
  the CO emission channel maps (see Fig.~\ref{co-channel}).
  The CO lines are very broad at the central position with velocities from
  $\sim-26.5$ to $\sim-23$\kms\ and from $\sim-20.5$ to $\sim-16$\kms, which
  indicates that the CO emission is probably tracing the outflowing gas there.

  Figure \ref{co-pvcuts}b shows the P-V cut obtained through the central
  position following an approximate East--West direction.
  From this P-V cut, the peak emission could be displaced $\sim2$--3 arcsec to
  the East.
  This position would coincide with the region where \hcop\ shows line
  broadening, East of SMM~1 (see Sect.~\ref{kinemhcop}).
  This P-V diagram also reveals two hyperbolic shapes at red- and blue-shifted
  channels, which could trace a possible keplerian rotation or infall.
  To explore this possibility, we tried to fit the centroid position of each
  velocity channel of the P-V plot with simulations of a keplerian rotating
  disk, within a range of central masses and inclinations.
  The best fits we found have $M/sin^2 i\sim15-20$\mo, and therefore seem to
  be incompatible with the mass determinations we have obtained from the dust
  continuum emission.

  There could also exist the possibility that the NE outflow and part of the
  ``E arm'' of the E--W outflow are related forming a wide opening outflow.
  The coincidence of the directions of the NE outflow and the strong optical
  jet do not seem to support this, and neither does the good agreement in the
  morphology and physical parameters (see Table~\ref{tbloutflows} and
  Sect.~\ref{outflowparams}) between the red- and blue-shifted arms of the
  E--W outflow.

  \citet{Magnier99} suspected ``Holoea'' to be an FU Orionis type star, which
  would explain the recorded increase in its luminosity in the last 40
  years.
  FU Orionis stars are variable stars \citep{Herbig77,HartmannKenyon96},
  thought to be low-mass young stellar objects closer to Class I protostars
  than to Class II T~Tauri stars \citep{Sandell01, Herbig03}, that experience
  a brightening of up to 6 mag over a few months, followed by a much slower
  fading back to their original luminosity, linked to eruptions caused by the
  increase of several orders of magnitude in the mass accretion rate through
  the circumstellar disk around the star \citep{HartmannKenyon96, Hartmann04,
    ReipurthAspin04}.
  A few FU Oris, such as V1057 Cyg \citep{Herbig03,ReipurthAspin04} and V1331
  Cyg \citep{McMuldroch93, Quanz07}, show a shell of material around them,
  possibly as a consequence of an energetic mass ejection event
  \citep{McMuldroch93}.
  We do not find these structures in our observations or clear kinematic
  indications of shells expanding through the molecular gas surrounding \iras.

 \begin{table*}

   \caption{
     Physical parameters of the main CO (1--0) outflows found around \iras
   }

   \centering

   \addtolength{\tabcolsep}{-1.3pt}

   \begin{tabular}{lccccccccccc}

     \hline\noalign{\smallskip}
 
     & $R_{\rm max}^a$ & $V_{\rm max}^b$ & area$^c$ & $S_{\nu}$ & $N_{CO}^d$ &
      $M_{\rm out}^e$ & $t_d^f$ & $\dot{M}_{\rm out}^g$ & $P_{\rm out}^h$ &
      $F_{\rm out}^i$ & $L_{\rm out}^j$\\
     Outflow & (pc) & (km\,s$^{-1}$) & (arcsec$^2$) & (Jy) & (cm$^{-2}$) &
      (\mo) & ($10^4$\,yr) & (\mo\,yr$^{-1}$) & (\mo\,km\,s$^{-1}$) &
      (\mo\,km\,s$^{-1}$\,yr$^{-1}$) & (\lo)\\

     \noalign{\smallskip}\hline\noalign{\smallskip}


     E--W red & 0.19~~ & 5.8~~ & 416 & 9.90 & $1.6\times10^{16}$ & 0.130~ &
      3.1 & $4.2\times10^{-6}$ & 0.76 & $2.4\times10^{-5}$ & $2.3\times10^{-2}$\\

     E--W blue & 0.15~~ & 4.1~~ & 592 & 13.7 & $9.9\times10^{15}$ & 0.117~ &
      3.6 & $3.2\times10^{-6}$ & 0.48 & $1.3\times10^{-5}$ & $8.4\times10^{-3}$\\

     NE red & 0.24~~ & 3.6~~ & 380 & 6.75 & $6.5\times10^{15}$ & 0.049~ & 6.7
      & $7.3\times10^{-7}$ & 0.17 & $2.6\times10^{-6}$ & $1.4\times10^{-3}$\\ 

     \noalign{\smallskip}\hline\noalign{\medskip}

   \end{tabular}

   \begin{minipage}{18cm}

      \begin{list}{}{\leftmargin 1.5em \rightmargin 1em}

      \addtolength{\itemsep}{-2pt}

       \item[$^a$] Projected maximum size of the outflow at the $3\sigma$
         contour, adopting a distance $D=1.2$~kpc.

       \item[$^b$] Maximum relative velocity of the outflow, either
         red-shifted or blue-shifted, with respect to the systemic velocity.

       \item[$^c$] area in the sky inside the $3\sigma$ contour of the
         integrated intensity maps over which the flux is measured.

       \item[$^d$] CO column density, calculated assuming local
         thermodynamical equilibrium (LTE) conditions.

       \item[$^e$] Mass traced by the outflow, calculated from
         $M_{\rm out} [M_{\odot}] = \bar{\mu} m_{\rm H} \Omega_{\rm S} D^2
         N_{\rm  CO}/X[{\rm CO}] $,  
         where $\bar{\mu}$ is the average molecular weight, 2.33, $m_{\rm H}$
         is the atomic hydrogen mass, $\Omega_{\rm S}$ is the solid angle of
         the source, $D$ is the distance to the source, and $X[{\rm CO}]$ is
         the adopted CO abundance relative to H$_2$.
         We used $X[{\rm CO}]=10^{-4}$ following \citet{Frerking82}.

       \item[$^f$] dynamical time scale of the outflow, $t_d = R_{\rm
         max}/V_{\rm max}$ 

       \item[$^g$] mass outflow rate, $\dot{M}_{\rm out} = M_{\rm out}/t_d$.

       \item[$^h$] momentum of the outflow, $P_{\rm out}= M_{\rm out} V_{\rm max}$.

       \item[$^i$] momentum flux, $F_{\rm out} = P_{\rm out}/t_d$.

       \item[$^j$] outflow mechanical luminosity, $L_{\rm out} = 0.5 M_{\rm
         out} V_{\rm max}^3/R_{\rm max}$.



     \end{list}

   \end{minipage}

 \label{tbloutflows}
 \end{table*}

\subsubsection{Mass and momentum of the CO outflows}
 \label{outflowparams}

  Table~\ref{tbloutflows} shows the estimated parameters of the two main
  outflows identified around \iras:
  the East-West outflow around the position of SMM~1 and the NE outflow from
  the position of SMM~2.
  We present the parameters of the red- and blue-shifted arms of the East-West
  outflow, and the parameters of the red-shifted emission of the material that
  we identified as belonging to the NE outflow.

  In order to calculate the mass of each outflow, we measured the flux inside
  the $3\sigma$ contour of the integrated intensity maps where we could find
  emission belonging to each outflow:
  from $-20.63$ to $-15.81$\kms\ for the red-shifted arm of the E--W outflow,
  from $-25.71$ to $-22.66$\kms\ for the blue-shifted arm of the E--W outflow,
  and from $-18.09$ to $-20.63$\kms\ for the NE outflow.
  We estimated the CO column densities traced by the red- and blue-shifted
  channels adopting $T_{\rm ex} = 20$~K \citep{Snell84,Lada85,Aso00}, assuming
  local thermodynamical equilibrium conditions (LTE) and optically thin
  emission.
  Given the assumption of optically thin emission, the calculated column
  density values should be taken as a lower limit to the column density,
  because the lines are probably optically thick.
  For instance, \citet{Bontemps96} adopted a mean factor of the opacity
  correction for the CO line of 3.5, following the detailed study of
  \citet{CabritBertout92}.
  We have also chosen not to use any inclination angle, $i$, for the
  calculation of dynamical time scale, the mass outflow rate, the momentum of
  the outflow and the momentum flux.
  In the case of correcting for the inclination angle, the linear size should
  be divided by $sin~i$, and the velocities by $cos~i$.
  \citet{Magnier96} estimated an inclination angle of $i=45\degr$ for the NE
  outflow, but we do not have any additional information about the orientation
  of the E--W outflow with respect to the line of sight.
  \citet{Bontemps96} estimate a mean correction factor of 2.9, corresponding
  to a random outflow orientation angle of 57.3$\arcdeg$ for the calculation
  of the CO momentum flux, $F_{\rm out}$.
  Thus, following \cite{Bontemps96} the corrected momentum flux of the
  outflows from Table \ref{tbloutflows}, should be $F_{\rm CO} = 2.9\times 3.5
  \times F_{\rm out} \sim 10\times F_{\rm out}$, taking into account the
  corrections from opacity and the inclination angle mentioned above.

  The resulting physical parameters (see Table~\ref{tbloutflows}) show that
  the E--W outflow appears to be more massive by a factor of 5, when adding
  the blue- and red-shifted arms, than the NE outflow.
  The dynamical time scale of the outflows is about a few times $10^4$ yr, but
  the E--W outflow would have a shorter dynamical time scale by a factor of
  $\sim 2$.
  The rest of the physical parameters also show that the E--W outflow is a
  more powerful outflow than the NE outflow.

 \begin{figure*}

   \centering

   \begin{minipage}{9.5cm}
     \includegraphics[width=\hsize]{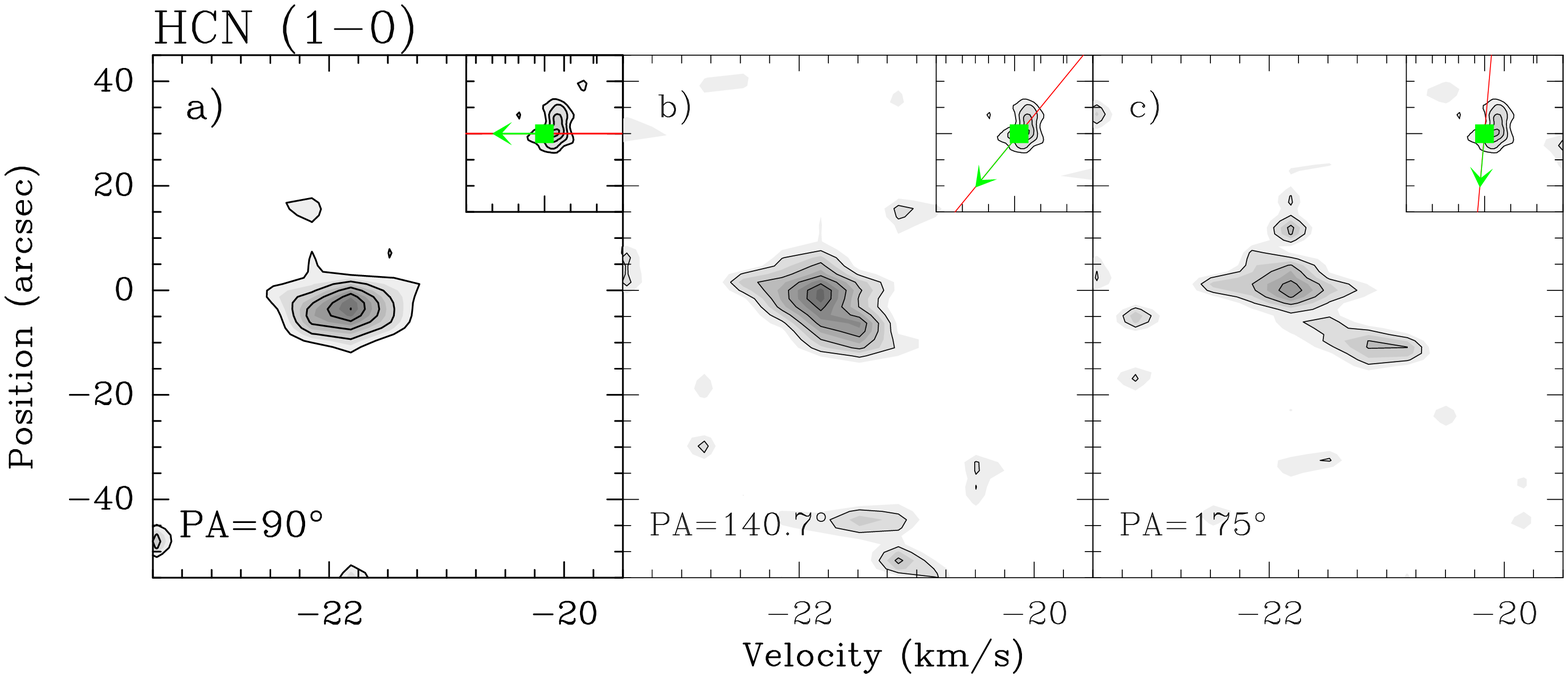}
     \includegraphics[width=\hsize]{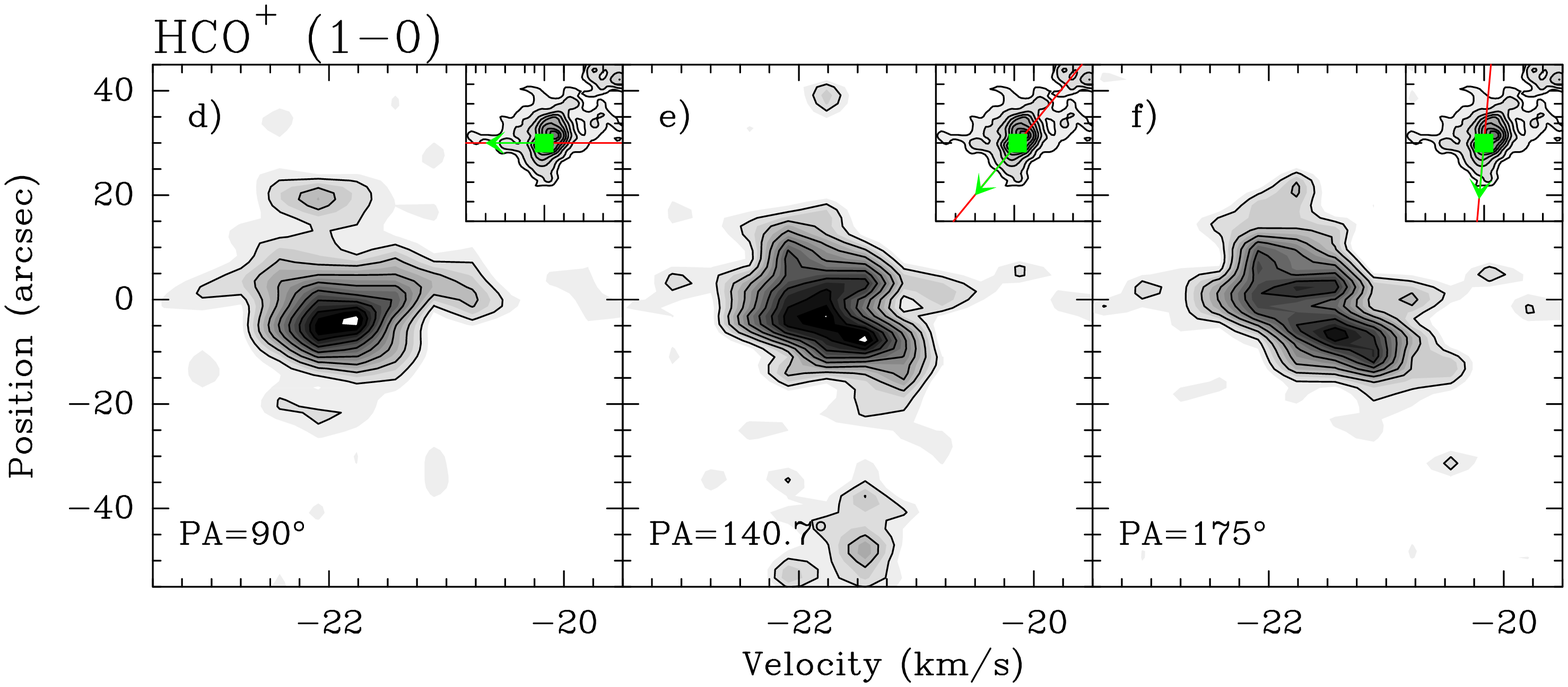}
   \end{minipage}
   \hspace{0.5mm}
   \begin{minipage}{8.2cm}
     \includegraphics[width=\hsize]{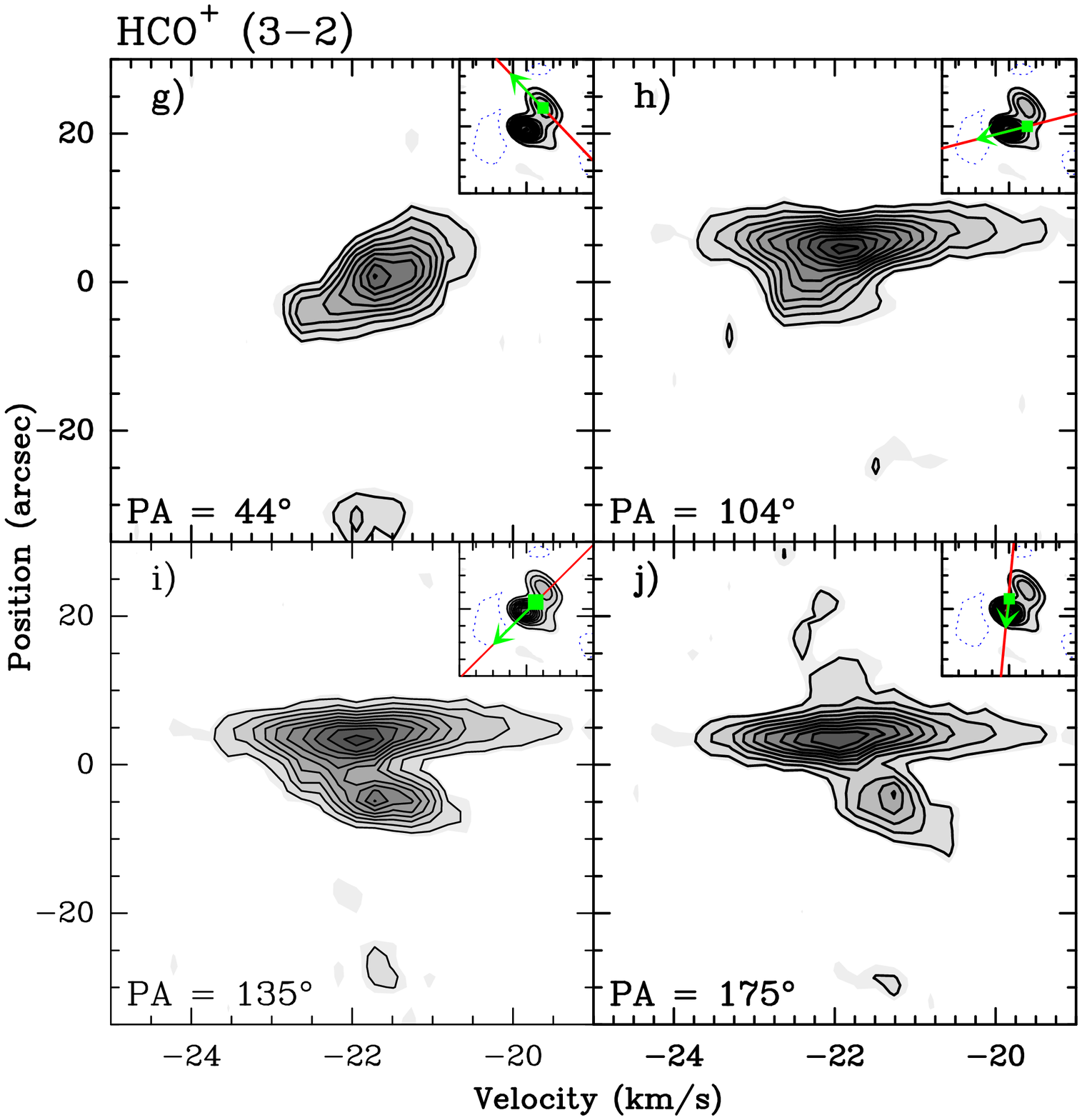}
   \end{minipage}

   \caption{
     \textit{Top left column)} Position-Velocity diagrams of the HCN (1--0)
     line along the directions:
     \textit{(a)} PA$=90\degr$, \textit{(b)} PA$=140.7\degr$ and \textit{(c)}
     PA$=175\degr$.
     The initial contour is at 0.30 \jyb, with increments of 0.20 \jyb.
     \textit{Bottom left column)} Position-Velocity diagrams of the
     \hcop\ (1--0) line along the directions:
     \textit{(d)} PA$=90\degr$, \textit{(e)} PA$=140.7\degr$, \textit{(f)}
     PA$=175\degr$.
     The initial contour is at 0.42 \jyb, with increments of 0.28 \jyb.
     \textit{Right column)} Position-Velocity diagrams of the \hcop\ (3--2)
     line along the directions:
     \textit{(g)} PA$=44\degr$, \textit{(h)} PA$=104\degr$, \textit{(i)}
     PA$=135\degr$, and \textit{(j)} PA$=175\degr$.
     The initial contour is at 0.72 \jyb, with increments of 0.72 \jyb.
     The inset maps follow the same conventions as in Fig.~\ref{co-pvcuts}.
   }

 \label{hcophcn-pvcuts}
 \end{figure*}

  We compared the momentum flux derived for our outflows with the momentum
  flux vs. radio continuum luminosity correlation of \citet{Anglada96}.
  We applied to $F_{\rm out}$ the correction factor of $\sim10$ used by
  \citet{Bontemps96}.
  \citet{AngladaRodriguez02} found a radio continuum luminosity of 0.22 mJy
  kpc$^2$ for their VLA~2 source.
  Unfortunately, the VLA observations were not able to resolve this source,
  but as we discussed in Sect.~\ref{sma_cont}, it is safe to assume that all
  the centimeter emission comes from SMM~1 and that the contribution coming
  from SMM~2 is much smaller.
  In this case, we find that the radio continuum luminosity of VLA~2 (SMM~1)
  falls very well in the correlation with the calculated momentum flux of the
  E--W outflow.
  For SMM~2, we can assume a $3\sigma$ upper limit for its associated radio
  centimeter continuum luminosity of $<0.1$ mJy kpc$^2$.
  Following the correlation of \citet{Anglada96}, the centimeter luminosity
  that we would obtain from the momentum flux of the NE outflow would be
  $\sim0.02$ mJy kpc$^2$, and thus consistent with the VLA upper limit.

  We also compared the physical parameters that we derived for our outflows
  with some of the results of \citet{Bontemps96}.
  First, we compared the corrected momentum flux of the outflows, $F_{\rm
    out}$, vs the bolometric luminosity, \lbol. 
  We find that the SMM 1 and the E--W outflow lie in the upper right corner of
  Fig.~5 of \citet{Bontemps96}, in the region associated with Class~0 sources,
  but relatively close to the best fit correlation found for Class~I sources.
  The location of SMM~2 and the NE outflow is more uncertain, but it would be
  located in the Class~I region, close to the best fit correlation for Class~I
  objects if \lbol$\sim2$--10\lo. 
  When we compared the momentum flux of the outflows with the mass of the
  circumstellar envelope derived in Sect.~\ref{dustemission} with Fig.~6 of
  \citet{Bontemps96}, we found that SMM~1 is also located in the Class~0 YSOs
  region, and close to the best fit correlation for the Class~I sources.
  On the other hand, SMM~2 and the NE outflow would be close to the region
  where there is the change from Class~0 to Class~I sources.

  Finally, we compared our data with the $F_{\rm out} c/$\lbol\ vs $M_{\rm
    env}/$\lbol$^{0.6}$ plot shown in Fig.~7 of \citet{Bontemps96} that tries
  to remove luminosity effects from the momentum flux and the envelope mass.
  The efficiency of our outflows, $F_{\rm out} c/$\lbol, is $\sim220$--420 for
  the E--W outflow, for SMM1 bolometric luminosities
  \lbol$\sim$82--41~\lo, and between $\sim600$--125 for the NE outflow
  for luminosities of SMM~2$\sim2$--10\lo.
  These values place SMM~1 in the region of the youngest Class~I sources, and
  SMM~2 just at the transition between Class~0 and Class~I YSOs. 
  The location of SMM~2 is again much more uncertain due to our lack of
  knowledge about its bolometric luminosity.

  We also compared the results we derived from the outflows around \iras\ with
  the sample of \citet{Aso00} of CO outflows in Orion.
  In a similar way, we find that the E--W outflow properties set the sub-mm
  source in the region occupied by what \citet{Aso00} called ``dark cloud
  Class~0'' objects in the envelope mass vs momentum flux and bolometric
  luminosity vs. momentum flux plots.
  The location of SMM~2, more difficult to determine, would seem to be more in
  according with what \cite{Aso00} called ``dark cloud Class~I'' objects.

 \begin{figure*}

   \centering

   \includegraphics[width=15cm]{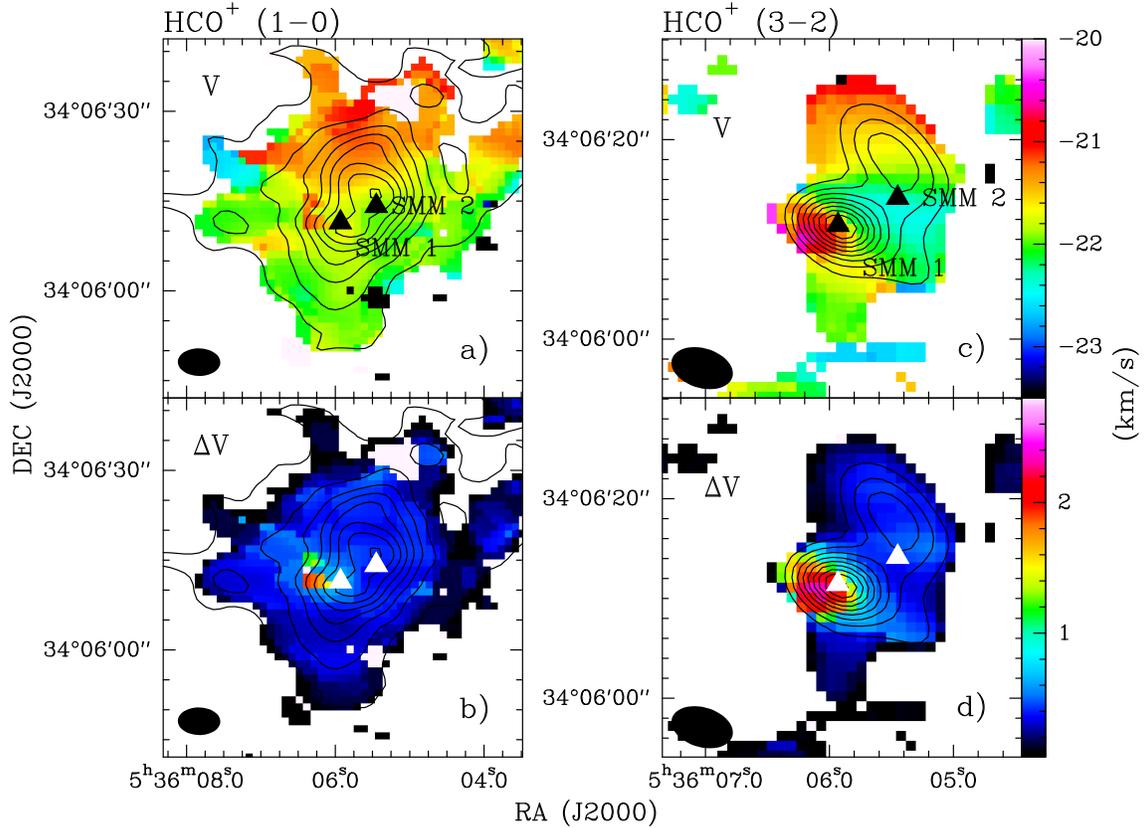}

   \caption{
     Maps of the first \textit{(top row)} and second order \textit{(bottom
       row)} moments of the \hcop\ (1--0) \textit{(left column)} and (3--2)
     \textit{(right column}) emission overlaid with the corresponding
     integrated intensity map in \iras. 
     The triangles mark the position of the continuum sources SMM~1 and SMM~2.
   }

 \label{hcop-moments}
 \end{figure*}

\subsection{Kinematic structures traced by \hcop\ and HCN}
 \label{kinemhcop}

  Figure~\ref{hcophcn-pvcuts} shows the Position-Velocity (P-V) cuts made for
  the HCN (1--0), \hcop\ (1--0), and \hcop\ (3--2) lines along some of the gas
  structures found in the integrated intensity maps.
  We find a velocity gradient in an approximately North--South direction,
  indicated by a velocity shift from $\sim-22.4$ to $-20.8$\kms\ for HCN and
  \hcop\ (1--0) in the P-V cuts following an almost N--S direction,
  PA$=175\arcdeg$ (Fig.~\ref{hcophcn-pvcuts}c and \ref{hcophcn-pvcuts}f).
  This velocity shift is also seen along the line connecting the two peaks in
  the integrated emission maps, PA$=140.7\arcdeg$,
  (Figs.~\ref{hcophcn-pvcuts}b and \ref{hcophcn-pvcuts}e), but with a
  shallower gradient.
  The first order moment map of the \hcop\ (1--0) line
  (Fig.~\ref{hcop-moments}a) clearly shows this velocity gradient as a
  transition from red- to blue-shifted velocities in an approximately
  North--South direction.
  The P-V cuts done for the \hcop\ (3--2) line show similar results, but in
  this case, the extension is considerably reduced:
  Figs.~\ref{hcophcn-pvcuts}i and \ref{hcophcn-pvcuts}j show the two
  condensations found in the integrated intensity maps and again the larger
  gradient is found for PA$\simeq175\arcdeg$ (Fig.~\ref{hcophcn-pvcuts}j).
  The first order moment map of the \hcop\ (3--2) emission
  (Fig.~\ref{hcop-moments}c) also shows the approximately North--South
  gradient found in \hcop\ (1--0) (Fig.~\ref{hcop-moments}a), but there is
  also a change in velocity East to South-West of SMM~1, that was not so
  clearly seen in the (1--0) line.

  The P-V cut taken in a SW--NE direction along the major axis of the NW
  condensation for the \hcop\ (3--2) line (Fig.~\ref{hcophcn-pvcuts}g),
  recovers the gradient shown in the \hcop\ (3--2) channel maps from
  $\sim-22.5$ to $\sim-21$\kms\ (Fig.~\ref{hcop32chan}).
  This P-V map shows a main condensation at a velocity $\sim-21.7$\kms,
  corresponding to the \hcop\ (3--2) NW peak, and two emission plateaus that
  trace other structures.
  The first one is located $\sim5$ arcsec to the NE at $\sim-21.1$\kms and it
  is also found in the P-V cut with PA$=175\arcdeg$
  (Fig.~\ref{hcophcn-pvcuts}j).
  The second one is located $\sim4$ arcsec to the SW, at $\sim-22.3$\kms, is
  also found in the P-V cut connecting SMM~1 and SMM~2 (PA$\simeq104\arcdeg$),
  and would correspond to SMM~2.  (Fig.~\ref{hcophcn-pvcuts}h).

  The \hcop\ (3--2) P-V cuts show much more clearly a line broadening of $\sim
  2$\kms\ at a position of $\sim2$--3 arcsec East of the the main
  \hcop\ emission peak.
  This is close to the position of the main CO (1--0) peak and the broadening
  could indicate the presence of some shocks at this position, but it could
  also be related to rotation or even collapse.
  This effect is outstanding in the second order maps of the \hcop\ lines
  (Fig.~\ref{hcop-moments}b and \ref{hcop-moments}d).
  \hcop\ line-widths are $\sim0.4$--0.6\kms\ for most of the map except for a
  region $6$ arcsec in size East of SMM~1.
  The \hcop\ (3--2) spectra on this part of the map show a complex structure,
  with shoulders on the line wings and maybe more than one component. 
  Line-widths can be as wide as 2.9\kms.
  If the line-widths corresponded to a rotational flow, we estimate,
  following \citet{Lada85}, that we would need a mass $M >8$\mo\ interior to
  the flow to support it, or $\sim25$\mo\ to have a bound self-gravitating
  structure, which is several times larger than the mass traced by the dust
  continuum.
  Therefore, the gas traced by \hcop\ close to SMM~1 is probably gas affected
  by the local outflows.
  We do not find any \hcop\ line broadening in the parts of the map to the
  North where there is red-shifted \hcop\ emission.

  Finally, the P-V cuts along a direction with PA$\simeq175\arcdeg$ of the
  \hcop\ and HCN lines (Figs.~\ref{hcophcn-pvcuts}c, \ref{hcophcn-pvcuts}f and
  \ref{hcophcn-pvcuts}j) also find traces of weak emission $\sim10$ arcsec
  South of SMM~1, which would correspond to the very weak condensation found
  in the respective channel maps.
  We also find weak emission $\sim20$ arcsec East of the central position of
  the map in the P-V cuts made in a direction East--West, PA$=90\arcdeg$,
  through the position of the \hcop\ peak (Figs.~\ref{hcophcn-pvcuts}d and
  \ref{hcophcn-pvcuts}a).
  This would correspond to the weak emission found in \hcop\ (1--0)
  coinciding with a local peak of CO (1--0).
  It is unclear what is the nature of these objects, probably gas in some
  stage of pre-stellar core or other clumpiness of the cloud.

\subsection{The distribution of the molecular gas}

\begin{figure}

   \centering

   \includegraphics[width=\hsize]{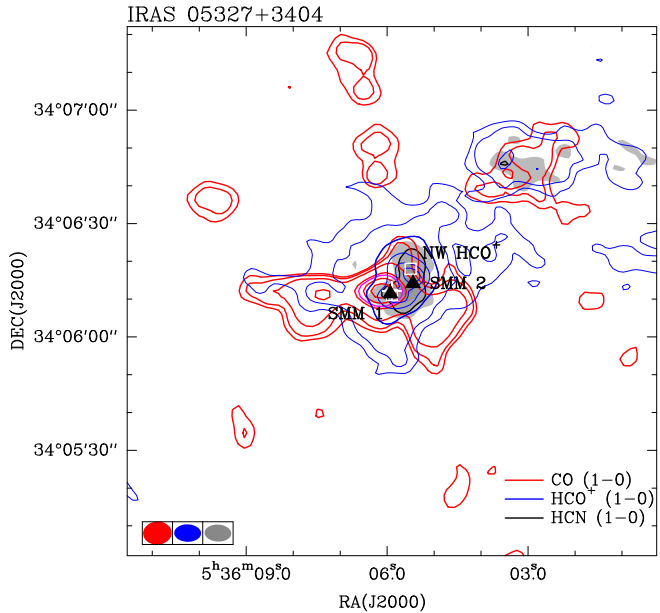}

   \caption{
     Integrated intensity emission of the CO (1--0) (red contours),
     \hcop\ (1--0) (blue contours), and HCN (1--0) (gray scale and one black
     contour at FWHM) emission, in the velocity intervals -27.47 to
     -15.04\kms, -22.42 to -20.78\kms, and -22.15 to -21.15\kms, respectively.
     The triangles mark the position of the continuum peaks and the
     squares the position of the \hcop\ (3--2) peaks.
  }

 \label{f-threelines}
 \end{figure}

\begin{figure}

   \centering

   \includegraphics[width=\hsize]{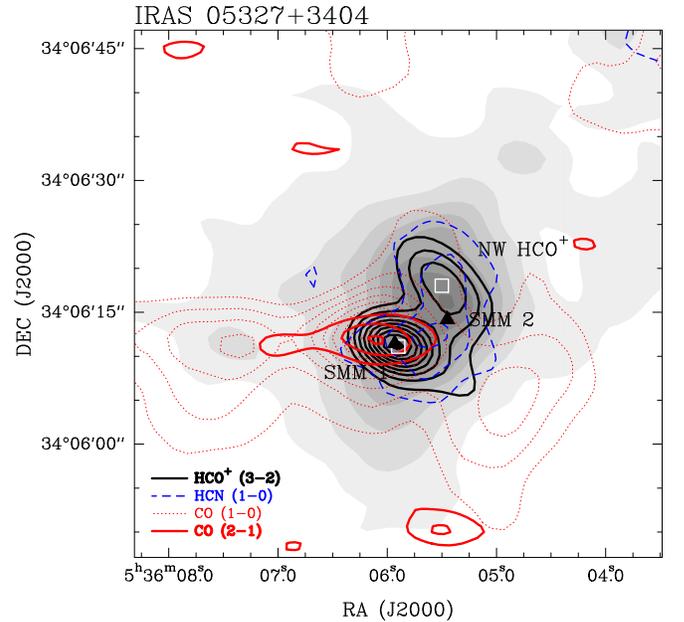}

   \caption{
     Integrated intensity emission of the \hcop\ (3--2) line \textit{(thick
       black contours)} overlaid over the integrated intensity of the CO
     (2--1) line \textit{(thick red contours)}, the integrated intensity
     map of \hcop\ (1--0) \textit{(gray scale)}, the integrated intensity map
     of HCN (1--0) \textit{(dashed blue contours)} and the integrated
     intensity map of CO (1--0) \textit{(dotted red contours)}.
     The triangles mark the position of the continuum peaks SMM~1 and
     SMM~2, and the squares, the position of the \hcop\ (3--2) peaks.
   }

 \label{f-smalines}
 \end{figure}

  Figure \ref{f-threelines} shows the superposition of the integrated
  intensity maps of the CO, \hcop, and HCN emissions obtained from the BIMA
  observation, mainly showing the gas distribution at larger scales.
  Figure \ref{f-smalines} shows a similar map zoomed in at smaller scales in
  order to compare the BIMA and SMA observations.

  The bulk of the more intense HCN and \hcop\ emission, which will trace the
  denser component of the molecular gas, is found in a central region,
  $\sim20\times15$ arcsec in size and it encompasses the sub-mm continuum
  emission and the two SMM sources, SMM 1 and 2.
  Lower density gas probably surrounds this central region, as it is traced by
  the weak \hcop\ (1--0) component.  
  The CO lines seem to trace mostly the molecular outflows.

  The molecular emission peaks show a large degree of coincidence amongst
  themselves and with the positions of the SMM sources:
  two of the \hcop\ peaks coincide with the SMM sources, while the third one
  (the one we named NW \hcop) also coincide with the northern HCN plateau.
  Only the HCN emission peak seems to be located in-between SMM 1 and SMM 2.
  Thus, the molecular and continuum data show that there are multiple objects
  in the region.
  Two of these objects are probably proto-stars, because of the detected
  sub-millimeter emission and its alignment with some of the high-velocity
  structures found in CO.
  The third condensation, only found in the \hcop\ and HCN lines, is either a
  starless or pre-stellar core; too cold to show a sub-millimeter peak.

  There are other minor emission peaks in our maps:
  a local CO (1--0) emission peak that coincides with a 6$\sigma$ peak of the
  \hcop\ (1--0) line at $\sim18$ arcsec East of \iras, and the hint of another
  possible condensation seen in the channel maps of \hcop\ (1--0) and (3--2)
  located $\sim10$ arcsec S of SMM~1.

\subsubsection{The missing CO emission}
 \label{missingco}

  There are probably different reasons to explain the big difference between
  CO and the other molecules, and why it is concentrated in a relatively
  East-West strip.
  CO traces lower density material than \hcop\ and HCN, and thus CO emission
  should be more widespread over the region.
  This makes it more likely to suffer from missing flux effects.
  From the low intensity distribution of the \hcop\ (1--0) emission, we would
  expect that CO would be distributed all around the molecular core, but most
  of the emission is probably filtered out by the interferometer.
  The integrated intensity map of CO mainly traces high-velocity gas
  (Fig.~\ref{co-channel}), which we expect would suffer less from
  filtering-out by the interferometer as it would be relatively more compact,
  and most of the CO emission we observe is probably due to molecular outflow
  emission.
  As a consequence (see Sect.~\ref{emco}), we do not find CO (1--0) emission
  around the systemic velocities almost anywhere in our map.  This prevents
  any chance of disentangling the relationship between the high-velocity and
  ``quiescent'' CO gas in the region.

\begin{figure}

   \centering

   \includegraphics[width=\hsize]{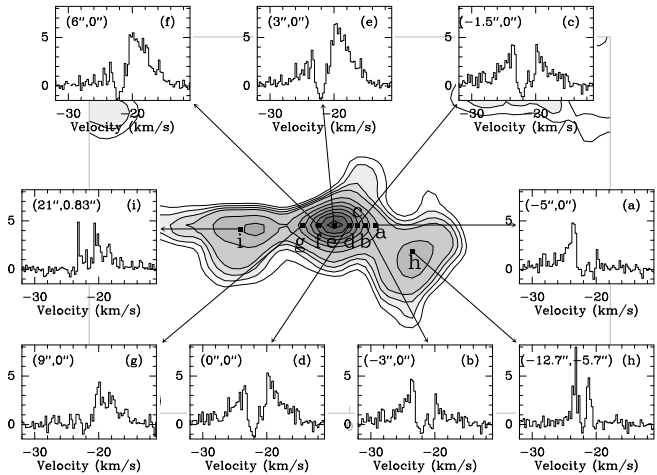}

   \caption{
     CO spectra, in units of \jyb, taken at nine different positions on both
     sides of the CO (1--0) emission peak of \iras.
     The offset positions are referenced to the nominal position of \iras.
   }

 \label{co-spectra}
 \end{figure}

  We took several spectra along the East--West direction on both sides of the
  CO integrated intensity maximum (Fig.~\ref{co-spectra}), in order to study
  the missing CO emission.
  The line emission in the spectra moves from mostly blue-shifted to the West
  of the emission peak (spectra ``a'' and ``b'') to mostly red-shifted to the
  East (spectra ``f'' and ``g''), with a gradual decrease of blue-shifted
  emission and an increase of red-shifted emission from the ``c'' to the ``e''
  spectra.
  All these spectra show missing CO emission around the systemic velocities
  ($-21.7$\kms), where we find what seems to be a more or less deep absorption
  feature in most of them.
  Only the farthest positions, ``a'' and ``i'', might not show this absorption
  feature.

  The spectral data could be indicating that there is an extended foreground,
  and colder, gas component surrounding the core revealed by \hcop\ and HCN,
  which is responsible for absorbing the CO emission.
  However, this absorption feature should also be apparent in single-dish
  observations, but that is not the case for the spectra of the CO (2--1) line
  of \citet{Magnier96}.
  Thus, we conclude that the interferometers mostly detect the CO components
  related to the outflows (relatively more compact and less subject to
  filtering out).

\subsection{The structure of the region and the nature of the powering
    sources}

  The millimetric and sub-millimetric observations show that the properties
  and the structure of the gas around the position of \iras\ ``Holoea''
  are much more complicated than previously thought.

  The data are helpful in the description of the morphology and the
  identification of existing proto-stellar objects, allow us to characterize
  several molecular outflows and argue for an identification of the respective
  powering sources.
  Nonetheless, our data cannot give any conclusive determination for several
  other velocity structures found in the region, and the possibility that
  there exist more embedded sources is still open.

  \citet{Magnier96} discussed three possible toy models that could describe
  the physical structure of M36 ``Holoea''.
  They favored an scenario in-between two of those models, in which there
  would be a molecular outflow powered by the star, with the red-shifted gas
  not moving in a well-collimated jet, but it would be spraying in a wide
  range of directions, or the presence of a thick disk around the star, \iras,
  would shadow the south-pointing jet and outflow and we would only see the
  south-side emission far from the star and at lower velocities.
  They also discussed the possibility that the system could be a binary
  \citep{Magnier99} in order to explain some of the properties of the observed
  SED.

  From the results of our data, we favor a different interpretation of the
  structure of the region.
  Our continuum data show that there are two sub-millimetric sources in the
  region, SMM~1 and SMM~2, separated by $\sim 6$ arcsec ($\sim7300$~AU), which
  would be powering the two main outflows we identify in the CO (1--0)
  observations:
  SMM~1 would be powering the East-West bipolar outflow, while SMM~2 would be
  the powering source of the NE red-shifted outflow.
  The orientation of both outflows would be different with respect to the line
  of sight, undetermined for the E--W outflow, while the NE outflow could have
  an inclination angle, $i=45\arcdeg$ \citep{Magnier99}.
  The blue-shifted counterpart of the NE outflow is not visible, maybe because
  its emission is confused with the blue-shifted lobe of the E--W outflow.
  Our data is not good enough to distinguish that, although this last
  possibility seems to be unlikely.

  The gas surrounding the two SMM objects would have masses from a few tenths
  of solar masses ($\sim0.8$\mo\ for SMM~2) up to 1--2~\mo\ ($\sim1.5$\mo\ for
  SMM~1).
  Additionally, the \hcop\ (1--0), \hcop\ (3--2), and HCN (1--0) lines also
  have emission peaks approximately coinciding with these two objects, which
  confirms that the SMM sources are associated with dense gas.
  The  comparison we  have been  able  to do  between the  \lbol\ measured  by
  \citet{Magnier96}, the envelope masses and the outflow parameters we derived
  for  our objects,  allows us  to guess  an evolutionary  stage for  both SMM
  sources.
  SMM~1 would be in a Class~0, probably very early Class~I, stage as indicated
  by the outflow parameters and the relationship between the bolometric
  luminosity and the envelope mass.
  SMM~2 seems to be in a more advanced stage of evolution, advanced Class~I
  probably, given the weaker molecular outflow and its outflow parameters.
  Additionally, SMM~1 seems to be related to a denser gas, traced by the
  continuum and high-density tracers, while SMM~2 seems to be in a slightly
  less dense environment.
  This also supports the earlier evolutionary stage for SMM~1.
  Thus, we think that our data rather shows that the \textit{two SMM sources
    are in an earlier evolutionary stage than previously expected, probably
    early Class~I stage}.
  We must be cautious though, given the uncertainties related to the
  calculation of the outflow parameters, the difficulties in the determination
  of the bolometric luminosity of each object, and the general complex
  kinematic structure of the region.

  There would be another object located close to the two SMM sources, $\sim
  8.5$ arcsec NW of SMM~2 (see Sect.~\ref{hcop32}), as detected by the
  \hcop\ and HCN lines.
  This object is not detected in the continuum, which suggests that it is
  probably a pre-stellar core or some cold embedded object.
  Interestingly, this core is also aligned with the direction of the NE
  outflow, but the lack of continuum emission associated with this object
  would a priori preclude it to be considered a candidate to be the powering
  source of any outflow.
  Additionally, there are two other condensations visible in some of the
  \hcop, HCN, and CO data, but not associated with continuum emission:
  one is located $\sim18$ arcsec East of SMM~1 and the other $\sim10$ arcsec
  South of SMM~1.
  With the limited amount of information that we have, these two condensations
  seem to be just two candidate starless cores.

  \citet{Magnier99} proposed that the system they found could be a binary,
  which is a possibility given the short projected distance between the SMM
  sources.
  Additionally, we also found a velocity gradient in the North--South
  direction, at at scale of $\sim 40$ arcsec, and another gradient in the gas
  around the NW \hcop\ position.
  These gradients suggest some internal motions, but there does not seem to be
  any significant relative motion between SMM~1 and SMM~2.

\section{Conclusions}
 \label{conclusions}

  We observed the object named ``Holoea'', associated with the \iras\ source
  in M36, with the SMA telescope at 1.1-mm, using the CO (2--1) and
  \hcop\ (3--2) lines, and with the BIMA telescope at 3-mm using the CO
  (1--0), \hcop\ (1--0), and HCN (1--0) lines.
  This object had been identified by \citet{Magnier96,Magnier99} as a
  candidate transitional YSO between Class~I and Class~II with several unusual
  properties.
  We detected the emission of the three molecules, but they show a markedly
  different distribution.
  We also detected dust continuum emission at 1.1-mm in the SMA observations.

  The continuum emission reveals two sub-millimeter peaks: the more intense
  one, SMM~1, with a flux density of 39.1 Jy, coincides within uncertainties
  with the nominal position of \iras\ and with the VLA~2 source of
  \citet{AngladaRodriguez02}, and we identify it as the sub-mm counterpart of
  the IRAS source.
  The second sub-mm emission peak, SMM 2, with a flux density of 19.9 Jy, is
  located $\sim6$ arcsec to the NW of SMM~1.
  From the dust emission, we calculated a mass of 1.5~\mo\ for SMM~1 and
  0.8~\mo\ for SMM~2.

  The distribution of the \hcop\ (1--0), (3--2), and HCN (1--0) emissions is
  more centrally concentrated, as it is expected for these molecular tracers,
  and they show at least two emission peaks.
  The main emission peak of the \hcop\ lines is found around the position of
  SMM 1, another \hcop\ (3--2) emission peak coincides with SMM 2, and there
  is a third peak, located 8.5 arcsec to the NW of SMM~1, not found in the
  continuum data, which is also present in the \hcop\ and HCN data.
  Thus, there is evidence of at least two proto-stellar objects and a third
  possible pre-stellar condensation.
  The integrated CO (1--0) emission is mainly found in an E--W direction, with
  the intensity peak $\sim3$ arcsec East of the position of SMM~1.
  The CO lines around the central position show a dip in the intensity, which
  we attribute to the filtering out of extended emission by the
  interferometers.

  We found several different velocity components in the red- and blue-shifted
  channels of the molecular lines.
  We identify two main molecular outflows in the region:
  a bipolar CO outflow elongated in an East--West direction with the blue- and
  red-shifted lobes overlapping over the position of SMM~1, which was also
  detected at lower angular resolution by \citet{Magnier99};
  and an outflow in the NE direction (PA$=9\arcdeg$), only found at
  red-shifted velocities, pointing to the position of SMM~2, in approximately
  the same direction as the outflow detected by \citet{Magnier99}.
  We identify SMM~1 as the best candidate to be the powering source of the
  E--W outflow, while SMM~2 could be the powering source of the NE outflow and
  the embedded YSO proposed by \citet{Magnier99}.
  We find other high-velocity structures around the central part of the map,
  and evidence of line broadening due to shocks and/or outflows in the CO and
  \hcop\ spectra, that might point to the presence of a third outflow or
  simply to the interaction of the E--W outflow with the surrounding gas.
  The sensitivity and angular resolution of our data does not allow us to
  avoid the confusion in the region.

  From the determination of the physical parameters of the two molecular
  outflows and the SMM objects, we have also been able to look into the
  evolutionary stage of the YSOs.
  We propose that SMM~1 is in an earlier evolutionary stage than SMM~2, and
  probably in an early Class~I phase.
  SMM~2 seems to be in a more advanced stage of the Class~I phase, but
  probably not by much.

  The general picture that we propose for the region is that it is a more
  complex system than it was thought, with at least two YSOs, which are
  powering the two main outflows detected in our CO channel maps.
  There could be more objects in the region, and additional outflows, but the
  detailed structure is very difficult to disentangle with the data available
  to us.
  Most of these objects are probably embedded, except ``Holoea'', which is
  beginning to be revealed, probably by the progressive clearing of the gas by
  one (or several) of the detected molecular outflows.
  Further multi-wavelength (optical, NIR, sub-millimetric) observations are
  needed to obtain a more precise description of the structure of the objects
  and of the physical processes that are taking part in them.

\acknowledgments

  We thank Alfonso Trejo for the help provided to retrieve the SMA filler-time
  data, and Glen Petitpas and Chunhua Qi for help in the reduction of the SMA
  data.
  We thank Ming-Fan Ho for his work on the BIMA data.
  O.~M.\ is supported by the NSC (Taiwan) ALMA-T grant to the Institute of
  Astronomy \& Astrophysics, Academia Sinica.
  The research of Y.-J.~K. was supported by NSC 99-2112-M-003-003-MY3 grant.

 {\it Facilities:} \facility{BIMA}, \facility{SMA}


\begin{thebibliography}{32}
\expandafter\ifx\csname natexlab\endcsname\relax\def\natexlab#1{#1}\fi

\bibitem[{Adams {et~al.}(1987)Adams, Lada, \& Shu}]{Adams87}
Adams, F.~C., Lada, C.~J., \& Shu, F.~H. 1987, ApJ, 312, 788

\bibitem[{Andr\'e {et~al.}(1993)Andr\'e, Ward-Thompson, \& M.}]{Andre93}
Andr\'e, P., Ward-Thompson, D., \& M., B. 1993, ApJ, 406, 122

\bibitem[{{Anglada}(1996)}]{Anglada96}
{Anglada}, G. 1996, in Astronomical Society of the Pacific Conference Series,
  Vol.~93, Radio Emission from the Stars and the Sun, ed. A.~R. {Taylor} \&
  J.~M. {Paredes}, 3--14

\bibitem[{Anglada \& Rodr\'{\i}guez(2002)}]{AngladaRodriguez02}
Anglada, G., \& Rodr\'{\i}guez, L.~F. 2002, RMxAA, 38, 13

\bibitem[{Aso {et~al.}(2000)Aso, Tatematsu, Sekimoto, Nakano, Umemoto, Koyama,
  \& Yamamoto}]{Aso00}
Aso, Y., Tatematsu, K., Sekimoto, Y., Nakano, T., Umemoto, T., Koyama, K., \&
  Yamamoto, S. 2000, ApJS, 131, 465

\bibitem[{{Blitz} {et~al.}(1982){Blitz}, {Fich}, \& {Stark}}]{Blitz82}
{Blitz}, L., {Fich}, M., \& {Stark}, A.~A. 1982, ApJS, 49, 183

\bibitem[{{Bontemps} {et~al.}(1996){Bontemps}, {Andre}, {Terebey}, \&
  {Cabrit}}]{Bontemps96}
{Bontemps}, S., {Andre}, P., {Terebey}, S., \& {Cabrit}, S. 1996, \aap, 311,
  858

\bibitem[{{Cabrit} \& {Bertout}(1992)}]{CabritBertout92}
{Cabrit}, S., \& {Bertout}, C. 1992, \aap, 261, 274

\bibitem[{Codella {et~al.}(1995)Codella, Palumbo, Pareschi, Scappini, Caselli,
  \& Attolini}]{Codella95}
Codella, C., Palumbo, G.~G.~C., Pareschi, G., Scappini, F., Caselli, P., \&
  Attolini, M.~R. 1995, MNRAS, 276, 57

\bibitem[{Frau {et~al.}(2010)Frau, Girart, Beltr\'an, Morata, Masqu\'e,
  Busquets, Alves, S\'anchez-Monge, Estalella, \& Franco}]{Frau10}
Frau, P., {et~al.} 2010, ApJ, 723, 1665

\bibitem[{Frerking {et~al.}(1982)Frerking, Langer, \& Wilson}]{Frerking82}
Frerking, M., Langer, W.~D., \& Wilson, R.~W. 1982, ApJ, 262, 590

\bibitem[{Hartmann {et~al.}(2004)Hartmann, Hinkle, \& Calvet}]{Hartmann04}
Hartmann, L., Hinkle, K., \& Calvet, N. 2004, ApJ, 609, 906

\bibitem[{Hartmann \& Kenyon(1996)}]{HartmannKenyon96}
Hartmann, L., \& Kenyon, S.~J. 1996, ARA\&A, 34, 207

\bibitem[{Herbig(1977)}]{Herbig77}
Herbig, G.~H. 1977, ApJ, 217, 693

\bibitem[{Herbig {et~al.}(2003)Herbig, Petrov, \& Duemmler}]{Herbig03}
Herbig, G.~H., Petrov, P.~P., \& Duemmler, R. 2003, ApJ, 595, 384

\bibitem[{{Ho} {et~al.}(2004){Ho}, {Moran}, \& {Lo}}]{hosma04}
{Ho}, P.~T.~P., {Moran}, J.~M., \& {Lo}, K.~Y. 2004, \apjl, 616, L1

\bibitem[{Hron(1987)}]{Hron87}
Hron, J. 1987, A\&A, 176, 34

\bibitem[{Lada(1985)}]{Lada85}
Lada, C.~J. 1985, ARA\&A, 23, 267

\bibitem[{Lada \& Wilking(1984)}]{LadaWilking84}
Lada, C.~J., \& Wilking, B.~A. 1984, ApJ, 287, 610

\bibitem[{Magnier {et~al.}(1999{\natexlab{a}})Magnier, Volp, Laan, van~den
  Ancker, \& Waters}]{Magnier99b}
Magnier, E.~A., Volp, A.~W., Laan, M.~E., van~den Ancker, M.~E., \& Waters, L.
  B. F.~M. 1999{\natexlab{a}}, A\&A, 352, 228

\bibitem[{Magnier {et~al.}(1999{\natexlab{b}})Magnier, Waters, Groot, van~den
  Ancker, Kuan, \& Mart\'{\i}n}]{Magnier99}
Magnier, E.~A., Waters, L. B. F.~M., Groot, P.~J., van~den Ancker, M.~E., Kuan,
  Y.-J., \& Mart\'{\i}n, E.~L. 1999{\natexlab{b}}, A\&A, 346, 441

\bibitem[{Magnier {et~al.}(1996)Magnier, Waters, Kuan, Chu, Taylor, Matthews,
  \& Mart\'{\i}n}]{Magnier96}
Magnier, E.~A., Waters, L. B. F.~M., Kuan, Y.-J., Chu, Y.-H., Taylor, A.~R.,
  Matthews, H.~E., \& Mart\'{\i}n, E.~L. 1996, {A\&A}, 305, 936

\bibitem[{McMuldroch {et~al.}(1993)McMuldroch, Sargent, \&
  Blake}]{McMuldroch93}
McMuldroch, S., Sargent, A.~I., \& Blake, G.~A. 1993, AJ, 106, 2477

\bibitem[{Ossenkopf \& Henning(1994)}]{Ossenkopf94}
Ossenkopf, V., \& Henning, T. 1994, A\&A, 291, 943

\bibitem[{Quanz {et~al.}(2007)Quanz, D., \& Henning}]{Quanz07}
Quanz, S.~P., D., A., \& Henning, T. 2007, ApJ, 656, 287

\bibitem[{Reipurth \& Aspin(2004)}]{ReipurthAspin04}
Reipurth, B., \& Aspin, C. 2004, ApJ, 608, L65

\bibitem[{Sandell \& Weintraub(2001)}]{Sandell01}
Sandell, G., \& Weintraub, D.~A. 2001, ApJSS, 134, 115

\bibitem[{Sault {et~al.}(1995)Sault, Teuben, \& Wright}]{Sault95}
Sault, R.~J., Teuben, P.~J., \& Wright, M.~C.~H. 1995, in Astronomical Data
  Analysis Software and Systems IV, ed. R.~A. Shaw, H.~E. Payne, and J.~J.
  Hayes (San Francisco: Astronomical Society of the Pacific), ASP Conf. Ser.,
  77, 433

\bibitem[{Snell {et~al.}(1984)Snell, Scoville, Sanders, \& Erickson}]{Snell84}
Snell, R.~L., Scoville, N.~Z., Sanders, D.~B., \& Erickson, N.~R. 1984, ApJ,
  284, 176

\bibitem[{{Sunada} {et~al.}(2007){Sunada}, {Nakazato}, {Ikeda}, {Hongo},
  {Kitamura}, \& {Yang}}]{Sunada07}
{Sunada}, K., {Nakazato}, T., {Ikeda}, N., {Hongo}, S., {Kitamura}, Y., \&
  {Yang}, J. 2007, \pasj, 59, 1185

\bibitem[{Ward-Thompson(1996)}]{WardThompson96}
Ward-Thompson, D. 1996, Ap\&SS, 239, 151

\bibitem[{Wouterloot {et~al.}(1993)Wouterloot, Brand, \& Fiegle}]{Wouterloot93}
Wouterloot, J.~G.~A., Brand, J., \& Fiegle, K. 1993, A\&ASS, 98, 589

\end{thebibliography}

\end{document}